\documentclass[sigconf, nonacm]{acmart}

\newcommand\vldbdoi{10.14778/3421424.3421431}
\newcommand\vldbpages{50 - 60}
\newcommand\vldbvolume{14}
\newcommand\vldbissue{1}
\newcommand\vldbyear{2021}
\newcommand\vldbauthors{\authors}
\newcommand\vldbtitle{\shorttitle} 
\newcommand\vldbavailabilityurl{https://github.com/megagonlabs/starmie}
\newcommand\vldbpagestyle{empty}

\usepackage[linesnumbered,ruled,vlined]{algorithm2e}

\usepackage{url}
\usepackage{verbatim}
\usepackage{graphicx}
\usepackage{multirow}
\usepackage{balance}
\usepackage{lscape}
\usepackage{float}
\usepackage{xcolor}
\usepackage{paralist}
\usepackage{booktabs}
\usepackage{xspace}
\usepackage{listings}
\usepackage{lipsum}
\usepackage{subcaption}

\definecolor{ForestGreen}{rgb}{0.0, 0.66, 0.47}
\definecolor{RubineRed}{rgb}{1.0, 0.0, 0.31}

\newcommand{\revision}[1]{{{{#1}}}}

\DeclareMathOperator*{\argmax}{\arg\!\max}

\usepackage{dsfont}

\newcommand{\name}{\ensuremath{\mathsf{Starmie}}\xspace}
\newcommand{\tus}{\textsf{TUS}\xspace}
\newcommand{\dtl}{\textsf{$D^3$L}\xspace}
\newcommand{\santos}{\textsf{SANTOS}\xspace}
\newcommand{\santosBench}{\textsf{SANTOS Small}\xspace}
\newcommand{\realBench}{\textsf{SANTOS Large}\xspace}
\newcommand{\tusSmallBench}{\textsf{TUS Small}\xspace}
\newcommand{\tusLargeBench}{\textsf{TUS Large}\xspace}
\newcommand{\sato}{\ensuremath{\mathsf{SATO}}\xspace}
\newcommand{\slk}{\ensuremath{\mathsf{Sherlock}}\xspace}
\newcommand{\single}{\ensuremath{\mathsf{SingleCol}}\xspace}

\newcommand{\bigA}{\mathcal{A}}
\newcommand{\bigC}{\mathcal{C}}
\newcommand{\bigF}{\mathcal{F}}
\newcommand{\bigH}{\mathcal{H}}
\newcommand{\bigM}{\mathcal{M}}
\newcommand{\bigO}{\mathcal{O}}
\newcommand{\bigT}{\mathcal{T}}
\newcommand{\bigS}{\mathcal{S}}

\begin{document}

\title{Semantics-aware Dataset Discovery from Data Lakes with Contextualized Column-based Representation Learning}

\author{Grace Fan}
\affiliation{%
  \institution{Northeastern University}
  \country{United States}
}
\email{fan.gr@northeastern.edu}

\author{Jin Wang}
\affiliation{%
  \institution{Megagon Labs}
  \country{United States}
}
\email{jin@megagon.ai}
\author{Yuliang Li}
\affiliation{%
  \institution{Megagon Labs}
  \country{United States}
}
\email{yuliang@megagon.ai}
\author{Dan Zhang}
\affiliation{%
  \institution{Megagon Labs}
  \country{United States}
}
\email{dan_z@megagon.ai}
\author{Renée Miller}
\affiliation{%
  \institution{Northeastern University}
  \country{United States}
}
\email{miller@northeastern.edu}

\begin{abstract}
	Dataset discovery from data lakes is essential in many real application scenarios.
    In this paper, we propose \name, an end-to-end framework for dataset discovery from data lakes (with table union search as the main use case).
	Our proposed framework features a contrastive learning method to train column encoders from pre-trained language models in a fully unsupervised manner.
    The column encoder of \name captures the rich contextual semantic information within tables by leveraging a contrastive multi-column pre-training strategy.
    We utilize the cosine similarity between column embedding vectors as the column unionability score and propose a filter-and-verification framework that allows exploring a variety of design choices to compute the unionability score between two tables accordingly.
	Empirical  results on real table benchmarks  show that \name outperforms the best-known solutions in the effectiveness of table union search by 6.8 in MAP and recall. 
	Moreover, \name is the first to employ the HNSW (Hierarchical Navigable Small World) index to accelerate query processing of table union search which provides a 3,000X performance gain over the linear scan baseline and a 400X performance gain over an LSH index (the state-of-the-art solution for data lake indexing).
\end{abstract}

\maketitle
\setcounter{page}{1}

\pagestyle{\vldbpagestyle}
\begingroup\small\noindent\raggedright\textbf{PVLDB Reference Format:}\\
\vldbauthors. \vldbtitle. PVLDB, \vldbvolume(\vldbissue): \vldbpages, \vldbyear.\\
\href{https://doi.org/\vldbdoi}{doi:\vldbdoi}
\endgroup
\begingroup
\renewcommand\thefootnote{}\footnote{\noindent
This work is licensed under the Creative Commons BY-NC-ND 4.0 International License. Visit \url{https://creativecommons.org/licenses/by-nc-nd/4.0/} to view a copy of this license. For any use beyond those covered by this license, obtain permission by emailing \href{mailto:info@vldb.org}{info@vldb.org}. Copyright is held by the owner/author(s). Publication rights licensed to the VLDB Endowment. \\
\raggedright Proceedings of the VLDB Endowment, Vol. \vldbvolume, No. \vldbissue\ %
ISSN 2150-8097. \\
\href{https://doi.org/\vldbdoi}{doi:\vldbdoi} \\
}\addtocounter{footnote}{-1}\endgroup

\vspace{-2mm}

\ifdefempty{\vldbavailabilityurl}{}{
\vspace{.3cm}
\begingroup\small\noindent\raggedright\textbf{PVLDB Artifact Availability:}\\
The source code, data, and/or other artifacts have been made available at \url{\vldbavailabilityurl}.
\endgroup
}

\section{Introduction}\label{sec:intro}

The growing number of open datasets from governments, academic institutions, and companies have brought new opportunities for innovation, economic growth, and societal benefits.
To integrate and analyze such datasets, researchers in both academia and industry have built a number of dataset search engines to support the application of dataset discovery~\cite{DBLP:conf/icde/FernandezAKYMS18,DBLP:journals/debu/MillerNZCPA18,DBLP:conf/www/BrickleyBN19,DBLP:journals/pvldb/CasteloRSBCF21,DBLP:conf/icde/SantosBMF22,DBLP:conf/cikm/GalhotraK20,DBLP:journals/pvldb/LimayeSC10}. 
One popular example is Google's dataset search~\cite{DBLP:conf/www/BrickleyBN19} which provides keyword search on the metadata.
However, for open datasets, simple keyword search might suffer from data quality issues of incomplete and inconsistent metadata across different datasets and publishers~\cite{DBLP:journals/pvldb/NargesianZPM18,DBLP:journals/pvldb/NargesianZMPA19,DBLP:journals/pvldb/AdelfioS13,DBLP:conf/sigmod/FaridRIHC16}.
Thus it is essential to support \emph{table search} over open datasets, and more generally data lake tables (including private enterprise data lakes), to boost dataset discovery applications, such as finding related tables, domain discovery, and column clustering.

Finding related tables from data lakes~\cite{DBLP:conf/sigmod/SarmaFGHLWXY12,DBLP:conf/sigmod/ZhangI20,DBLP:journals/pvldb/Miller18} has a wide spectrum of real application scenarios.
There are two sub-tasks of finding related tables, namely table union search and joinable table search.
In this paper, we mainly focus on the problem of \emph{table union search}, which has been recognized as a crucial task in dataset discovery from data lakes~\cite{DBLP:conf/icde/BogatuFP020,DBLP:conf/sigmod/ZhangI20,DBLP:journals/pvldb/NargesianZMPA19,DBLP:journals/pvldb/ZhuNPM16,DBLP:journals/pvldb/NargesianZPM18,santos23,DBLP:journals/pvldb/Miller18}.
Given a query table and a collection of data lake tables, table union search aims to find all tables that are unionable with the query table.
To determine whether two tables are unionable, existing solutions first identify all pairs of unionable columns from the two tables based on  column representations, such as bag of tokens or bag of word embeddings.
They then devise some mechanism to aggregate the column-level results to compute the table unionability score.

\revision{{\bf State-of-the-art:} Early work on finding unionable tables used table clustering followed by simple syntactic measures such as the difference in column mean string length and cosine similarities to determine if two tables are unionable~\cite{DBLP:journals/pvldb/CafarellaHK09}.  Table union search~\cite{DBLP:journals/pvldb/NargesianZPM18} improved on this by applying a rich collection of column representations including syntactic, semantic (leveraging ontologies), and natural language (based on word-embeddings) column representations. Two important innovations of this work were the modeling of data lake context to create an {\em ensemble unionability score} which models the surprisingness of a score given the score distributions within a data lake and the use of LSH indices to make table union search fast over large data lakes~\cite{DBLP:journals/pvldb/NargesianZPM18}.  More recently $D^3L$~\cite{DBLP:conf/icde/BogatuFP020} added additional column representations based on regular expression matching and SANTOS~\cite{santos23} added to the column representations, representations of binary relationships.  In parallel to these search-based approaches, the mighty hammer of deep learning has been applied to the problem of column matching (determining the semantic type of a column)~\cite{DBLP:journals/pvldb/ZhangSLHDT20,DBLP:conf/kdd/HulsebosHBZSKDH19}.  Since these approaches are supervised, they can only be applied to finding a limited set of semantic types (78 in their experiments), and while not a general solution for unionability in data lakes, they can be used in an offline fashion to find unionable tables containing the types on which they are trained.  
}

However, there are still plenty of opportunities to further improve the performance of table union search.
One important issue is to learn sufficient contextual information between columns in tables so as to determine the unionability.
This point can be illustrated in the following motivation example. 

\begin{figure}[!t]
	\centering
	\includegraphics[width=0.5\textwidth]{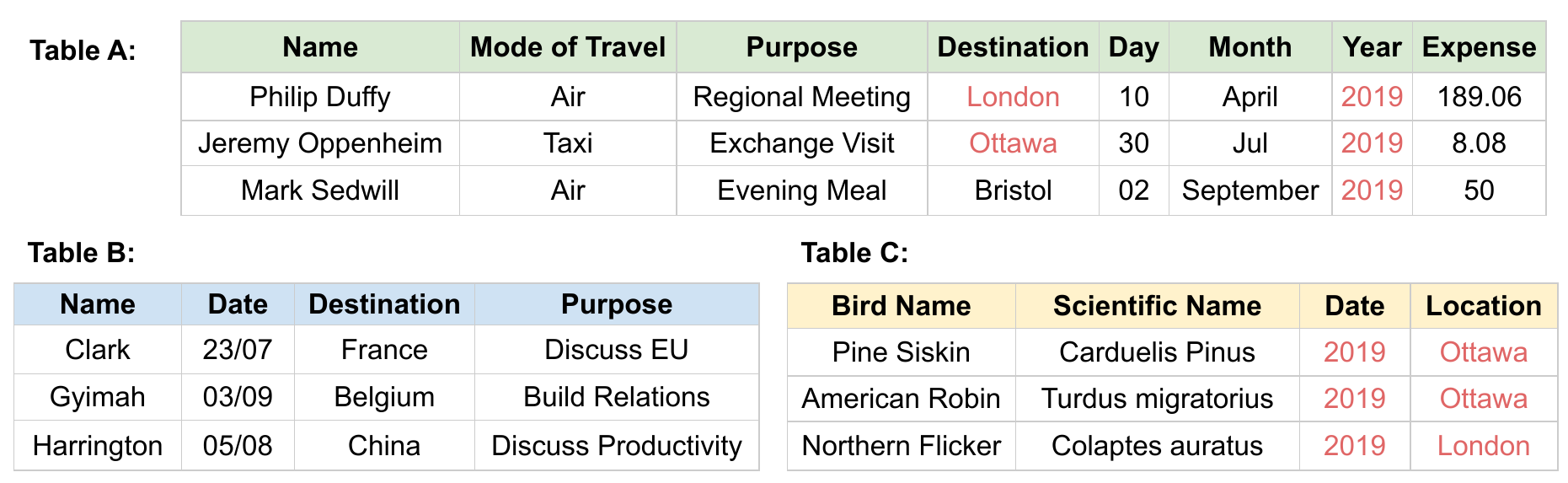}
	\vspace{-4mm}
	\caption{An example of table union search on Open Data.}\vspace{-1em}
	\label{fig:motivation}
	\vspace{-2mm}
\end{figure}

\begin{example}
    Figure~\ref{fig:motivation} shows an example of finding unionable tables.
    Given the query Table A,  existing approaches first find unionable columns.
    In this example, the column {\tt Destination} in Table A will be deemed more unionable with {\tt Location} from Table C than with {\tt Destination} from Table B. 
    This is because the syntactic similarity score, e.g. overlap and containment Jaccard, between the two {\tt Destination} columns is 0; while the average word embedding of cities (Table A) is also not as close to that of nations (Table B).  Similarly, if an ontology is used, Table A and Table C shares the same class while the values in B are in different (though related) classes.
    Meanwhile, looking at the tables as a whole we observe that Table A is actually irrelevant to Table C. 
    But as existing solutions only look at the pair of single columns when calculating column unionability score, the columns {\tt Year/Date} and {\tt Destination/Location} of the two tables might be wrongly aligned together. 
    Even techniques that look at relationships~\cite{santos23} can be fooled by the value overlap in this relationship and determine the relationship {\tt Year-Destination} in Table A to be unionable with {\tt Date-Location} in Table C.
    This kind of mistake can be avoided by looking at \revision{a table's context, i.e. information carried by other columns within a table. Looking at the table as a whole, a method should be able to}  recognize that the {\tt Year} in Table A is part of a travel date while in Table C it is the date of discovery of a bird; and {\tt Destination} in Table A refers to the cities to which the officers are traveling; whereas {\tt Location} in Table C is the city where a bird is found.
\end{example}

From the above example, we focus on the following challenges in proposing a new solution.
Firstly, it is essential to learn richer semantics of columns based on natural language domain.
To this end, we require a more powerful approach to learn the column representation so as to capture richer information instead of relying on simple methods like the average over bag of word embeddings utilized in previous studies~\cite{DBLP:conf/icde/BogatuFP020,DBLP:conf/icde/DongT0O21} or even the similarity of the word embedding distributions~\cite{DBLP:journals/pvldb/NargesianZPM18}. 
Secondly, we argue that it is crucial to utilize the contextual information within a table to learn the representation of each column, which is ignored by previous studies. Even proposals for capturing relationship semantics do not use contextual information to learn column representations~\cite{santos23}.
Finally, due to the large volume of data lake tables, it is also a great challenge to develop a scalable and memory-efficient solution.

We propose \name, an end-to-end framework for dataset discovery from data lakes with table union search as the main use case.
\name uses pre-trained language models (LMs) such as BERT~\cite{DBLP:conf/naacl/DevlinCLT19} to obtain semantics-aware representations for columns of data lake tables.
While pre-trained LMs have been shown to achieve state-of-the-art results in table understanding applications~\cite{DBLP:journals/pvldb/0001LSDT20,DBLP:journals/pvldb/DengSL0020,DBLP:conf/sigmod/SuharaL0ZDCT22}, their good performance heavily relies
on high-quality labeled training data.
For the problem setting of table union search~\cite{DBLP:journals/pvldb/NargesianZPM18,DBLP:journals/pvldb/NargesianZMPA19}, we must come up with a fully unsupervised approach in order to apply pre-trained LMs to such applications, something not yet supported by previous studies.
\name addresses this issue by leveraging \emph{contrastive representation learning}~\cite{DBLP:conf/icml/ChenK0H20} to learn column representations in a {\bf self-supervised manner}.  \revision{An innovation of this approach is to assume that two randomly selected columns in a data lake can be used as negative training examples.  For positive examples, we propose and use novel data augmentation methods.}
The framework defines a learning objective that connects the same or similar columns in the representation space while separating distinct columns. 
As such, \name can apply the pre-trained representation model in downstream tasks such as table union search without requiring any labels. 
We also propose to combine the learning algorithm with a novel multi-column table transformer model to learn \emph{contextualized} column embeddings \revision{that model the column semantics depending on not only the column values, but also their context within a table}.
While a recent study \santos~\cite{santos23} can reach a similar goal by employing a knowledge base, our proposed methods can automatically capture such contextual information from tables in an unsupervised manner without relying on any external knowledge or labels.

Based on the proposed column encoders, we use cosine similarity between column embeddings as the column unionability score and develop a bipartite matching based method to calculate the table unionability score. 
We propose a filter-and-verification framework that enables the use of different indexing and pruning techniques 
to reduce the number of computations of the expensive bipartite matching.
While most previous studies employed LSH index to improve the search performance, we also make use of HNSW \revision{(Hierarchical Navigable Small World)} index~\cite{DBLP:journals/pami/MalkovY20} 
to accelerate query processing. 
Experimental results show that HNSW can significantly improve the query time while only slightly reducing the MAP/recall scores.
Besides table union search, we further conduct two case studies to show that \name can also support other dataset discovery applications such as joinable table search and column clustering.  \revision{We believe these results show great promise in the use of contextualized, self-supervised embeddings for many table understanding tasks.}

Our contributions can be summarized as the following.

\begin{compactitem}
	\item We propose \name, an end-to-end framework to support dataset discovery over data lakes with table union search as the main use case.
	\item We develop a contrastive learning framework to learn contextualized column representations for data lake tables without requiring labeled training instances. \name achieves an improvement of 6.8\% in both MAP and recall compared with \revision{the best state-of-the-art method, with a MAP of 99\%, a significant margin compared with previous studies}.
	\item We design and implement a filter-and-verification based framework for computing the table-level unionability score which can accommodate multiple design choices of indexing and pruning to accelerate the overall query processing. 
	By leveraging the HNSW index, \name achieves up to three orders of magnitude in performance gain for query time relative to the linear scan baseline.
	\item We conduct an extensive set of experiments over two real world data lake corpora.
	Experimental results demonstrate that the proposed \name framework significantly outperforms existing solutions in effectiveness.
	It also shows good scalability and memory efficiency.
	\item We further conduct case studies to show the flexibility and generality of our proposed framework in other dataset discovery applications. 
\end{compactitem}


\section{Overview}\label{sec:architecture}

\subsection{Problem definition}\label{subsec-problem}

A data lake consists of a collection of tables $\bigT$.
Each table $T \in \bigT$ consists of several columns $\{t_1, \dots, t_m\}$ where each column $t_i$ can be from different domains. 
Here $m$ is the number of columns in table $T$ (denoted as $|T| = m$).
We will use the notation $T$ to denote both the table and its set of columns if there is no ambiguity.
To determine the unionability between two columns, following previous studies, we employ \emph{column encoders} to generate the representations of columns.
Then the \emph{column unionability score} can be computed to measure the relevance between those representations.
A column encoder $\bigM$ takes a column $t$ as input and outputs $\bigM(t)$ as the representation.
Given two columns $t_i$ and $t_j$, the column unionability score is computed as $\bigF (\bigM(t_i), \bigM(t_j))$, 
where $\bigF$ is a scoring function between two column representations.


Based on the column unionability scores, we  compute the table unionability score between two tables, which is obtained by aggregating the column unionability scores introduced above. 
Given two tables $S$ and $T$, we define a table unionability scoring \revision{mechanism} as $U = \{\bigF, \bigM, \bigA\}$, where $\bigM$ and $\bigF$ are the column encoder and scoring function for two column representations, respectively.
Here $\bigA$ is a mechanism to aggregate the column unionability scores between all pairs of columns from the two tables.
We will introduce the details of $\bigA$ later in Section~\ref{sec:online}.

Following the above discussions, we can formally define the table union search problem as a top-k search problem as Definition~\ref{def-tus}:

\begin{definition}[Table Union Search] \label{def-tus}
	Given a collection of data lake tables $\bigT$  and a query table $S$, top-k table union search aims at finding a subset $\bigS \subseteq \bigT$ where $|\bigS| = k$ and $\forall T \in \bigS$ and $T' \in \bigT- \bigS $, we have $U(S, T) \geq U(S, T')$.
\end{definition}

\subsection{System architecture}\label{subsec-overview}

\begin{figure}[!h]
	\centering
	\includegraphics[width=0.48\textwidth]{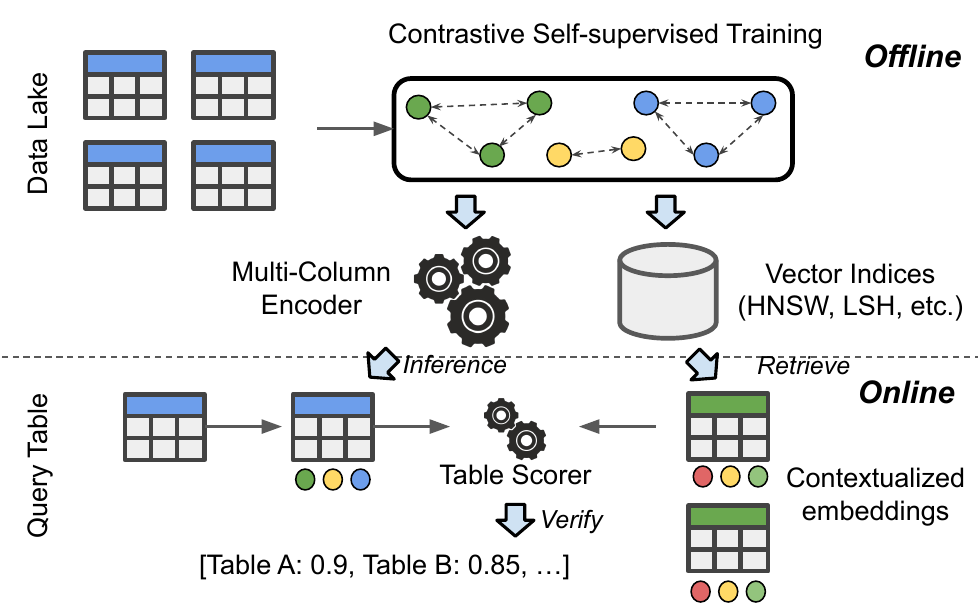}
	\caption{\small During the offline phase, \name pre-trains
	a multi-column table encoder using contrastive learning and stores the embeddings 
	of data lake columns in vector indices like HNSW. During online processing, \name 
	retrieves candidate tables with similar contextualized column embeddings then verifies their
	table-level unionability scores using column alignment algorithms.}\vspace{-2em}
	\label{fig:overall}
\end{figure}

Figure~\ref{fig:overall} shows the overall architecture of \name that solves table union search in two stages: offline and online.

During the offline stage, \name pre-trains a column representation model that encodes columns
of data lake tables into dense high-dimensional vectors (i.e., column embeddings).
Then, we apply the trained model to all data lake tables to obtain the column embeddings via model inference.
We store the embedding vectors in efficient vector indices for online retrieval.
A key challenge for the offline stage is to train high-quality column encoders that capture  the semantics of tabular data.
In \name, we follow a recent trend \cite{DBLP:journals/pvldb/0001LSDT20,DBLP:journals/pvldb/DengSL0020,DBLP:conf/sigmod/SuharaL0ZDCT22} of 
table representation learning that encodes tabular data using pre-trained language models (LMs). 
Pre-trained LMs have achieved state-of-the-art performance on table understanding tasks 
such as column type and relation type annotation~\cite{DBLP:conf/sigmod/SuharaL0ZDCT22}.
However, the good performance of pre-trained LMs requires fine-tuning on high-quality labeled datasets, which are always not available in table search applications such as table union search. 
Using  pre-trained LMs off-the-shelf is also problematic as the column embeddings cannot capture (ir-)relevance between columns or the contextual information within tables.
To this end, in Section~\ref{sec:offline}, we propose a contrastive learning framework for learning high-dimensional column representations in fully unsupervised manner. 
We combine the framework with a multi-column table model that captures column semantics from the column values while taking the table context into account.
Then we  apply the column encoder to all tables 
to convert each table into a collection of embedding vectors. 

During the online stage, given an input query table, we retrieve a set of candidate tables from the vector indices by searching for data lake column embeddings of high column-level similarity with the input columns.
\name then applies a verification step for checking and ranking the candidates for the top-$k$ tables with the highest table-level unionability scores. 
The first challenge for the online stage is how to efficiently search for unionable columns. 
This is not a trivial task due to the massive size of data lakes.
We address this challenge by allowing different design choices of state-of-the-art high-dimensional vector indices. 
Yet another challenge is designing a table unionability function that can effectively aggregate the column unionability scores. 
\revision{As in other studies, we employ weighted bipartite graph matching. 
To address its limitation of high computation complexity, we introduce a novel algorithm to reduce the number of expensive calls to the exact matching algorithm by deducing lower and upper bounds of the matching score (Section~\ref{sec:online}).}

\newcommand{\emb}{\mathsf{emb}}

\section{Learning contextualized column embeddings}\label{sec:offline}

We now describe the offline stage for training high-quality column encoders.
The encoder pre-processes tables into sequenced inputs and uses
a pre-trained LM to encode each column into a high-dimensional vector.
\revision{
We first introduce background knowledge in Section~\ref{subsec:bg}.
We describe a novel contrastive learning approach for table encoders in Section \ref{subsec:simclr} and generalize it to multi-column encoders for contextualized embeddings in Section \ref{subsec:contextualized}.
Finally, we describe the table pre-processing approaches to generate the input for such learning processes in Section~\ref{subsec:pretbl}.
}

\subsection{Background}
\label{subsec:bg}

\begin{figure}
	\centering
	\includegraphics[width=0.48\textwidth]{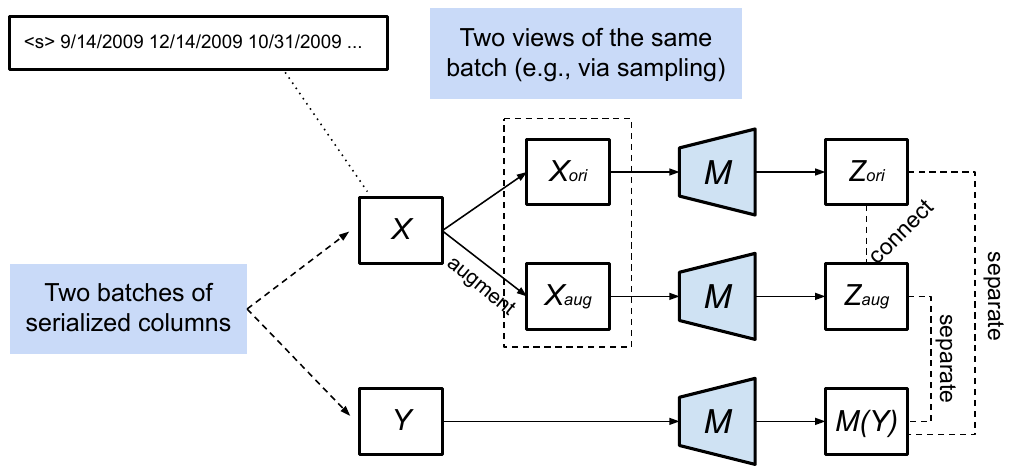}
	\caption{Contrastive learning with single-column input.}\vspace{-2mm}
	\label{fig:simclr}
\end{figure}

\revision{
Contrastive learning is a self-supervision approach that
learns data representations where similar data items are close while distinct data items are far apart. 
In \name, we adopt SimCLR~\cite{DBLP:conf/icml/ChenK0H20}
which was recently shown to be effective in Vision and NLP applications. 
Figure \ref{fig:simclr} illustrates the high-level idea of the algorithm.
The goal is to learn an encoder $\bigM$ (e.g., a column encoder)
that takes a data item (e.g., a column) as input and encodes it into a high-dimensional vector.
To train the encoder in a self-supervised manner without labels, SimCLR
relies on (1) a data augmentation operator generating semantic-preserving views
(in our context this means $X_{\mathsf{ori}}$ and $X_{\mathsf{aug}}$ that are unionable)
of the same data item and (2) a sampling method (e.g., uniform sampling from a large collection) 
that returns pairs of data items 
(i.e., $X$ and $Y$) that are distinct (meaning non-unionable) with high probability.
SimCLR then applies a contrastive loss function that connects the representations of the 
semantic-preserving (unionable) views meanwhile separating those of the sampled distinct (non-unionable) items.
Next, we illustrate how we apply the algorithm for training a single-column encoder.}



\subsection{Contrastive Learning Framework}
\label{subsec:simclr}


The 
goal is to connect representations of the same or unionable columns in their representation space while separating representations of distinct columns. 
To achieve the first goal, \revision{Algorithm~\ref{alg:simclr}} leverages a data augmentation operator $\mathsf{op}$ \revision{(Line~\ref{line:aug})}. 
Given a batch of columns $X = \{x_1, \dots, x_N\}$ where $N$ is the batch size, 
$\mathsf{op}$ transforms $X$ into a semantics-preserving view $X_{\mathsf{aug}}$. 
\revision{We design the augmentation operator to be uniform sampling of the values from the original column. By doing so, we can generate diverse views of the same column while all views preserve
the original semantic types. }
Then $\bigM$ can encode the batches $X$ (also $X_{\mathsf{ori}}$ which is a copy of $X$ in the figure) 
and $X_{\mathsf{aug}}$ into column embedding vectors $\vec{Z}_{\mathsf{ori}}$ and $\vec{Z}_{\mathsf{aug}}$ respectively.
Note that $\vec{Z}_{\mathsf{ori}}$ and $\vec{Z}_{\mathsf{aug}}$ are both matrices with size $N$ times
the dimension of embedding vector (e.g., 768 for BERT). 
\ref{line:aug}

Next, the algorithm leverages a \emph{contrastive loss function} to connect the semantics-preserving views of columns and separate representations of distinct columns \revision{(Line~\ref{line:encode})}. 
More specifically, let $\vec{Z} = \{\vec{z}_i\}_{1 \leq i \leq 2N}$ be the concatenation of the two encoded views $\vec{Z}_{\mathsf{ori}}$ and $\vec{Z}_{\mathsf{aug}}$ of batch $X$ introduced above. 
Here $\vec{z}_i$ is the $i$-th element of $\vec{Z}_{\mathsf{ori}}$ for $i \leq N$ and the $(i-N)$-th element of $\vec{Z}_{\mathsf{aug}}$ for $i > N$.
We first define a single-pair loss $\ell(i, j)$ for an element pair $(\vec{z}_i, \vec{z}_j)$ to be Equation~\ref{equ:infonce1}.
\begin{equation}\label{equ:infonce1}
\ell(i, j)=-\log \frac{\exp \left(\operatorname{sim}\left(\vec{z}_{i}, \vec{z}_{j}\right) / \tau\right)}{\sum_{k=1}^{2 N} \mathds{1}_{[k \neq i, \revision{k \neq j}]} \exp \left(\operatorname{sim}\left(\vec{z}_{i}, \vec{z}_{k}\right) / \tau\right)}
\end{equation}
where \emph{$\operatorname{sim}$} is a similarity function such as cosine and $\tau$ is a temperature hyper-parameter in the range $(0, 1]$.
We fix $\tau$ to be 0.07 empirically.
Intuitively, by minimizing this loss for a pair $(\vec{z}_i, \vec{z}_j)$ that are views of the same columns, we 
(i) maximize the similarity score $\operatorname{sim}\left(\vec{z}_{i}, \vec{z}_{j}\right)$ in the numerator and 
(ii) minimize $\vec{z}_i$'s similarities with all the other elements in the denominator. 



Next, we can obtain the contrastive loss by averaging all matching pairs shown in Equation~\ref{equ:infonce2} \revision{(Line~\ref{line:contrast})}:
\begin{equation}\label{equ:infonce2}
\mathcal{L}_{\mathsf{contrast}} =\frac{1}{2 N} \sum_{k=1}^{N}[\ell(k, k+N)+\ell(k+N, k)] 
\end{equation}
where each term $\ell(k, k+N)$ and $\ell(k+N, k)$ refers to pairs of views generated from the same column.


\SetKwInOut{Parameter}{Variables}
\begin{algorithm}[!t]
\small
	\KwIn{ A collection $D$ of data lake columns}
	\Parameter{ Number of training epochs $\mathsf{n\_epoch}$; \\
	       Data augmentation operator $\mathsf{op}$; Learning rate $\eta$ }
	\KwOut{ An embedding model $\bigM$
	        }
	Initialize $\bigM$ using a pre-trained LM\;
	\For{ $\mathsf{ep} = 1$ to $\mathsf{n\_epoch}$}{
	    Randomly split $D$ into batches $\{B_1, \dots B_n\}$\;
	    \For{$B \in \{B_1, \dots B_n\}$} {
	        \tcc{augment and encode every item}
	        $B_{\mathsf{ori}}, B_{\mathsf{aug}} \leftarrow \mathsf{augment}(B, \mathsf{op})$\;
	        \label{line:aug}
	        $\vec{Z}_{\mathsf{ori}}, \vec{Z}_{\mathsf{aug}}  \leftarrow \bigM (B_{\mathsf{ori}}), \bigM(B_{\mathsf{aug}})$\;
	        \label{line:encode}
	        \tcc{Equation (\ref{equ:infonce1}) and (\ref{equ:infonce2})}
	        $\mathcal{L} \leftarrow \mathcal{L}_{\mathsf{contrast}}(\vec{Z}_{\mathsf{ori}}, \vec{Z}_{\mathsf{aug}})$\;
	        \label{line:contrast}
	        \tcc{Back-prop to update $\bigM$}
	        $\bigM \leftarrow \textsf{back-propagate}(\bigM, \eta, \partial \mathcal{L} / \partial \bigM) $\;
	    }
	}
	\Return $\bigM$\;
	\caption{$\textsf{SimCLR pre-training}$}
	\label{alg:simclr}
\end{algorithm}

\subsection{Multi-column Table Encoder}
\label{subsec:contextualized}

While the method shown in Algorithm~\ref{alg:simclr} learns column representations based on values within a column itself, it cannot take the contextual information of a table into account.
For example, the single-column model can understand that a column consisting of values ``1997 1998 \dots''
is a column about years, but depending on the context of other columns present in the same table, the same column can represent 
``years in which a species of bird was observed in a specific area'' or ``years of car production'', etc.
As illustrated in the example in Figure~\ref{fig:motivation}, such understanding is important for deciding whether two tables are unionable or not.

\begin{figure}[!t]
    \centering
    \includegraphics[width=0.48\textwidth]{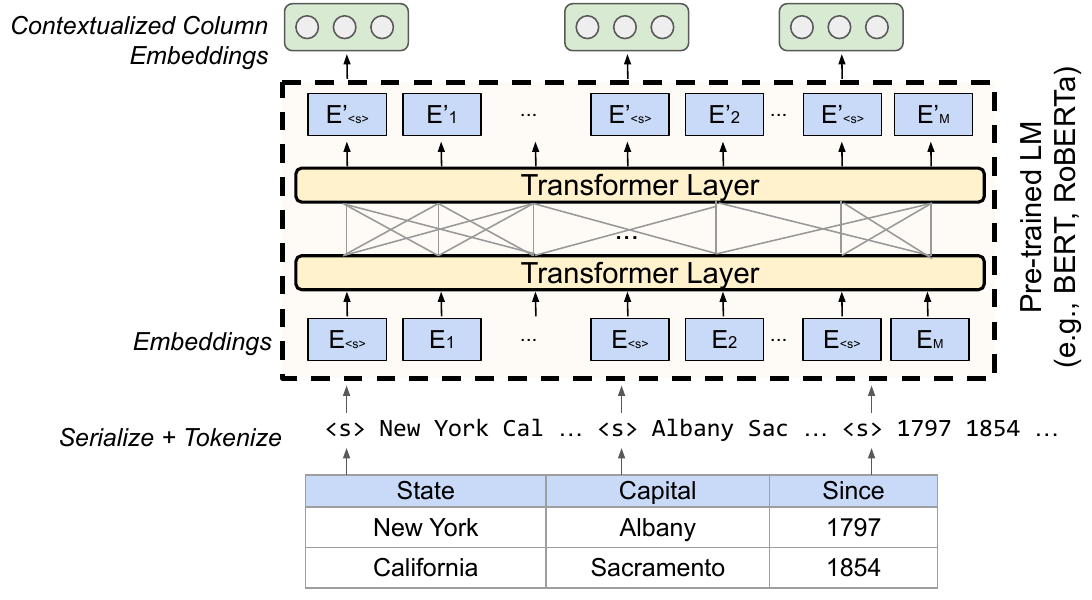}
    \vspace{-2em}
    \caption{Multi-column table encoder.}\vspace{-1em}
    \label{fig:lm}
\end{figure}

To address this problem, \name combines contrastive learning with a \emph{multi-column table encoder} illustrated in Figure~\ref{fig:lm}. 
The model starts with serializing an input table into a string by concatenating cell values from each column.
\revision{Following the implementation of tokenizers in the HuggingFace library, it also adds} a special separator token ``\textsf{<s>}'' to indicate the start of each column. 
Next, we feed the sequence as the input to a pre-trained LM such as RoBERTa~\cite{DBLP:journals/corr/abs-1907-11692}. \revision{Since the special token ``\textsf{<s>}'' at the start
a sequence in RoBERTa is pre-trained to capture the sequence representations, 
we also expect it to capture representations of columns given the table context.}

The pre-trained LM first converts the input sequence into a sequence of token embeddings independent
of their context then applies 12 or more Transformer layers~\cite{DBLP:conf/nips/VaswaniSPUJGKP17} on top.
The self-attention mechanism in the Transformer layers convert the word embeddings into a sequence of \emph{contextualized embeddings}. 
These vector representations depend not only on the tokens themselves (e.g., ``1797'') but also their context (e.g., ``Albany''). 
As such, we can extract the  representations of the separator tokens (i.e., ``\textsf{<s>}'') to be the contextualized column embeddings.

\begin{figure}[!t]
    \centering
    \includegraphics[width=0.4\textwidth]{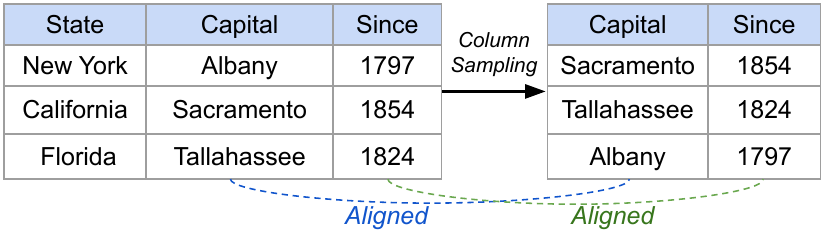}
    \caption{Table-level augmentation and column alignment.}
    \label{fig:column_sampling}\vspace{-1em}
\end{figure}

To apply contrastive learning using the multi-column model, we adapt the SimCLR
algorithm (Algorithm \ref{alg:simclr}) as follows.
First, we create the batches of columns (Line 3) by uniformly sampling a batch of tables from all data lake
tables and form each batch of columns $B$ using all columns from the sampled tables.
To augment the batch $B$, instead of transforming each column independently, we apply \emph{table-level}
augmentation operators such as row sampling and column sampling (Line 5). 
Note that in the multi-column setting, the augmentation operators produce views of tables with pairs of columns that align 
with each other. 
These pairs form the positive pairs in the contrastive loss as we illustrate in Figure~\ref{fig:column_sampling}.


We summarize the supported augmentation operators in Table \ref{tab:da}. 
While there is a large design space of the operators, we summarize them
by the levels (e.g., cell, row, column) of the table to which the operators apply.
\revision{The cell-level operators are general transformations also used in related
tasks such as Entity Matching~\cite{DBLP:journals/pvldb/0001LSDT20}.
The row and column-level operators cover  different ways for creating samples of rows/columns.}
One can also perform more complex transformations by applying multiple operators simultaneously.
\revision{In our ablation study (see Appendix B.1), we find that the simple 
column sampling operator (drop\_col) provides the best performance.} 

\begin{table}[!b]
\vspace{-2mm}
\caption{Data augmentation operators at different levels.}\label{tab:da}\vspace{-1em}
\small
\begin{tabular}{lp{2.8cm}p{4cm}}\toprule
Level  & Operators                                           & Description                                                            \\ \midrule
Cell   & drop\_cell, drop\_token, swap\_token, repl\_token    & Dropping a random cell; Dropping/swapping tokens within cells          \\ \midrule
Row    & sample\_row,  \mbox{shuffle\_row}                   & Sampling x\% (e.g., 50) of rows; Shuffling the row order               \\ \midrule
Col & drop\_col, drop\_num\_col, shuffle\_col & Dropping$X$  (numeric) columns; Shuffling  column order \\ \bottomrule
\end{tabular}
\end{table}

\begin{figure}[!t]
    \centering
    \includegraphics[width=0.48\textwidth]{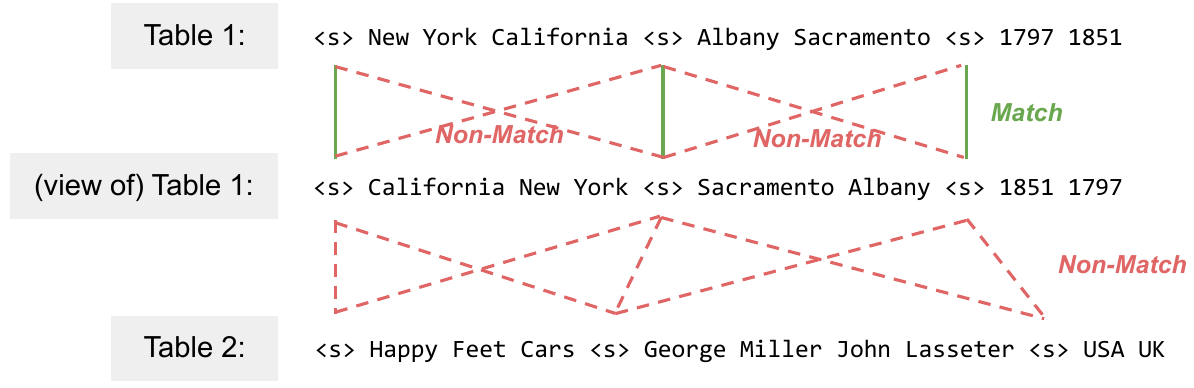}
    \vspace{-2mm}
    \caption{Contrastive learning positive and negative pairs.}
    \label{fig:match_nonmatch}\vspace{-3mm}
\end{figure}

We then apply the multi-column model on the original and augmented views of tables to obtain the contextualized 
column embeddings $\vec{Z}_{\mathsf{ori}}$ and $\vec{Z}_{\mathsf{aug}}$ (Line 6) and compute the contrastive loss (Line 7).
Note that in the multi-column setting, the positive pairs 
(for which we maximize the similarity)
consist of the aligned pairs of columns generated by the augmentation operators. 
We minimize the similarity of all other pairs which include (i) pairs of unaligned columns from the same table and 
(ii) all pairs of columns from two distinct tables. 
By doing so, the algorithm learns representations that can distinguish columns with the same/different table contexts, thus creating the positive and negative pairs shown in Figure~\ref{fig:match_nonmatch}.
More formally, let $P$ be the set of indices of all aligned pairs of columns in the batch $B$, we minimize the multi-column contrastive loss shown in Equation~\ref{equ:infonce3}:
\begin{equation}\label{equ:infonce3}
\mathcal{L}_{\mathsf{multi\text{-}column}} =\frac{1}{2|P|} \sum_{(i, j)\in P}[\ell(i, j)+\ell(j, i)] .
\vspace{-2mm}
\end{equation}

\subsection{Table Preprocessing }
\label{subsec:pretbl}

Typical pre-trained LMs like BERT support an input length of at most 512 sub-word tokens, while a column in real-world tables such as those in Open Data may contain thousands or even millions of tokens.
\revision{
To apply the proposed techniques in Section~\ref{subsec:simclr} and~\ref{subsec:contextualized} on data lake tables, we must preprocess the columns to reduce the input length to fit
the token limit of LMs, while preserving their semantics.
The procedure is outlined in Algorithm \ref{alg:pretable}, while
the full details with design choices 
(scoring functions, row/column orders, and alignment rules) are in the appendix due to the space limitation.

\SetKwInOut{Parameter}{Variables}
\begin{algorithm}[!t]
	\small
	\KwIn{A table $T$; A token scoring function such as TF-IDF $\mathsf{TF\text{-}IDF(\cdot)}$; 
              The max \#tokens $m$.}
        \Parameter{Preprocessing $\mathsf{mode} \in \text{\{``row'', ``cell'', ``token''\}}$}
	\KwOut{The table $T'$ with selected rows, cells, or tokens 
	}
        \ForEach{cell $c \in T$}{
            \tcc{Sum over token scores}
            $\mathsf{cell\_score(c)} \leftarrow \sum_{\text{token } t \in c} \mathsf{TF\text{-}IDF(c)}$\;
        }
 
        \ForEach{row $r \in T$}{
            \tcc{Sum over cell scores}
            $\mathsf{row\_score(c)} \leftarrow \sum_{\text{cell }c \in r} \mathsf{cell\_score(c)}$\;
        }
 
        \If {$\mathsf{mode} = \text{``row''}$}{
            \Return Top-$n$ rows with highest $\mathsf{row\_score}$ up to length $m$\;
        }
        \If {$\mathsf{mode} = \text{``cell''}$}{
            \Return Top-$n$ cells with highest $\mathsf{cell\_score}$ for each column
            up to length $m / |T|$\tcp*{$|T|$: number of columns}
        }
        \If {$\mathsf{mode} = \text{``token''}$}{
            \Return Top-$n$ tokens with highest $\mathsf{TF\text{-}IDF}$ for each column
            up to length $m / |T|$\tcp*{$|T|$: number of columns}
        }
	\caption{Table Preprocessing}
	\label{alg:pretable}
\end{algorithm}	

Algorithm~\ref{alg:pretable} illustrates the steps of table pre-processing.
It first assigns an importance score for each cell by first computing the TF-IDF scores of every token in a cell and then averaging the TF-IDF scores of all tokens.
Then it ranks the average cell-level scores of rows and then selects the rows to be included in the serialization result.
Here we finish this step in a deterministic way: by ranking in the descending order of the importance score, until we reach the token budget for each column.
}

\section{Online query processing}\label{sec:online}

In this section, we introduce how to find unionable tables based on contextualized column embeddings.
We first \revision{introduce} the table unionability scores and the overall workflow of online query processing in Section~\ref{subsec-tblunion}.
Then we discuss the design choices for reducing the number of candidates using vector indices 
and deducing bounds for more efficient verification in Sections~\ref{subsec-index} and~\ref{subsec-bound}, respectively.
Note that the online processing techniques explored here are not limited to any specific column encoders, they are also applicable to other dense-vector column representation methods~\cite{DBLP:conf/kdd/HulsebosHBZSKDH19,DBLP:journals/pvldb/ZhangSLHDT20}.

\subsection{Table-level Matching Score} \label{subsec-tblunion}

After training a column encoder $\bigM$ using techniques from Section~\ref{sec:offline}, we can then obtain the embedding vectors for all columns in data lake tables via model inference. 
The column unionability score between two columns $s$ and $t$ can be calculated by using cosine similarity as $\bigF$ between those embedding vectors. 
Next, we define the function $\bigA$ for aggregating the column unionability scores to compute the table unionability.
Motivated by the idea of $c$-alignment~\cite{DBLP:journals/pvldb/NargesianZPM18} that aims to find a maximum set of one-to-one alignment between columns in two tables, we propose modeling table unionability as a \emph{weighted bipartite graph matching} problem.
More formally, given two tables $S$ and $T$ with $m$ and $n$ columns respectively, we construct a bipartite graph $G = \langle S, T ,E\rangle$ where the nodes $S$ and $T$ are the two sets of columns.
The edges in $E$ denote the column unionability score between each pair of columns. 
Then table unionability score $U(S,T)$ can be calculated by finding the maximum bipartite matching of graph $G$.
In order to remove the noise caused by dissimilar pairs of columns, we follow the de-noising strategy from fuzzy string matching~\cite{DBLP:conf/icde/WangLZ19} by introducing a hyper-parameter $\tau$ as the 
similarity lower bound: given two columns $s \in S$ and $t \in T$, there is an edge $\langle s, t\rangle \in E$ iff $\bigF(s, t) \geq \tau$.  

\begin{figure}[!t]
	\centering
	\includegraphics[width=0.4\textwidth]{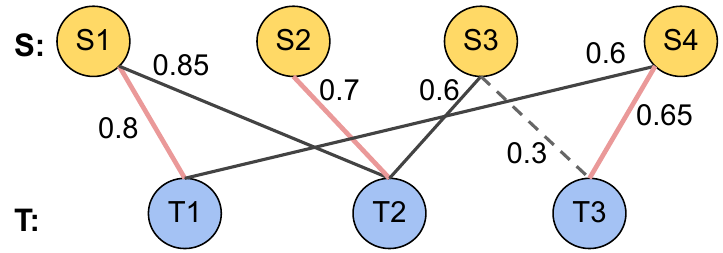}
	\caption{Example of table unionability score via maximum bipartite matching. Solid (red) lines denote the edges belonging to the maximum matching.}\vspace{-2mm}
	\label{fig:bp}
\end{figure}

\begin{example}
	We show an example of computing the table unionability score in Figure~\ref{fig:bp}.
	Suppose there are two tables $S$ and $T$ with 4 and 3 columns respectively and the threshold $\tau$ for column unionablity score is 0.5. 
	Since the cosine similarity between $s_3$ and $t_3$ is 0.3 ($<\tau$), the edge between them is discarded (denoted with a dash line).
	For the ease of presentation, we omit the remaining dash lines between other nodes in the figure. 
	The maximum bipartite matching of this graph consists of the edges in red (solid lines), which are $\langle s_1, t_1 \rangle$, $\langle s_2, t_2 \rangle$ and $\langle s_4, t_3 \rangle$ with a score of 2.15. 
\end{example}

In order to find the tables with top-k highest table unionability scores with the given query table $S$, a straightforward method is to conduct a linear scan: we use a min-heap with cardinality of $k$ to keep the results of top-k search, then for each table $T$ in the data lake, we directly compute $U(S,T)$; and if the score is higher than the top element of the min-heap, we replace the top element with it and adjust the min-heap accordingly. 
However, since the time complexity of weighted bipartite matching is $\bigO(n^3 \log n)$, where $n$ is the total number of columns in two tables, it is rather expensive to traverse all tables in a data lake.
A scalable solution requires reducing (i) the number of accessed tables and (ii) the computational overhead of verifying each pair of tables.

\SetKwInOut{Parameter}{Variables}
\begin{algorithm}[!t]
	\small
	\KwIn{$S$: the query table; $\bigT$: the set of data lake tables;}
	\Parameter{
		$k$: the number of desired results; \\
		$\tau$: threshold of column unionable score; \\
	}
	\KwOut{ $\bigH$: The top-k unionable tables}
	Initialize $\bigH$ and $\bigC$ as $\emptyset$;\\
	\For {all columns $s \in S$}{
		$\bigC = \bigC \cup \mathsf{findCandidates}(s, \tau, \bigT)$;
	}
	\For {all tables $T \in \bigC$}{
		\uIf{$|\bigH| < k$}{ 
			Compute $\mathsf{Verify}(S, T)$ and add $T$ into $\bigH$;
		}
		\Else{
			$X \leftarrow$ the score of top element of $\bigH$;\\
			\uIf{$\textbf{LB}(S,T) > X$}{
				Replace the top element of $\bigH$ with $T$;
			}
			\ElseIf{$\textbf{UB}(S,T) \leq X$}{
				Discard $T$;
			}
			\ElseIf{$\mathsf{Verify}(S, T) > X$}{
				Replace the top element of $\bigH$ with $T$;
			}
		}
	}
	\Return $\bigH$;
	\caption{\textsf{Online Query Processing}}
	\label{alg:topk}
\end{algorithm}	

We propose a filter-and-verification framework to address this issue as illustrated in Algorithm~\ref{alg:topk}.
Instead of doing a linear scan over all data lake tables, it employs filter mechanisms to identify a set of candidate tables $\bigC$ for further verification  (line:~3).
As a result, it can reduce the number of expensive verification operations $\mathsf{Verify}(S, T)$.
This is realized by the function \textsf{findCandidates} (Section~\ref{subsec-index}).
Then for all the candidate tables, we further come up with a pruning mechanism to estimate the lower bound $\textbf{LB}(S,T)$ and upper bound $\textbf{UB}(S,T)$ of $U(S,T)$. 
If the lower bound is larger than the current lowest score, we can directly replace it with the top element without further verification (line:~10).
Similarly, if the upper bound is no larger than the current lowest score, we can directly discard it (line:~12).
This pruning mechanism is effective since $\textbf{LB}$ and $\textbf{UB}$ are much more efficient to estimate than the exact verification $\mathsf{Verify}(S, T)$ (Section~\ref{subsec-bound}).

\subsection{Reducing the Number of Candidates}\label{subsec-index}

Given a column with its embedding vector, we need to quickly identify tables from the data lake that contain unionable columns, which is realized by the \textsf{findCandidates} function in Algorithm~\ref{alg:topk}.
This is a problem of similarity search over high-dimensional vectors.
Locality Sensitivity Hashing (LSH)~\cite{DBLP:conf/vldb/GionisIM99} has been used in previous studies of table search to find joinable~\cite{DBLP:journals/pvldb/ZhuNPM16}, unionable~\cite{DBLP:journals/pvldb/NargesianZPM18}, and related columns~\cite{DBLP:conf/icde/BogatuFP020} in sub-linear time.
The basic idea is to use a family of hash functions to map high-dimensional vectors into a number of buckets, where the probability that two vectors are hashed into the same bucket is correlated to the value of a certain similarity metric between them.
Following this work, we build a \textsf{simHash}~\cite{DBLP:conf/stoc/Charikar02} LSH index to estimate the cosine similarity between column embedding vectors.
Then for each query column vector $s$, we can quickly find a set of similar column vectors via an index lookup.
Then the candidate set $\bigC$ can be obtained by the union of candidates returned by utilizing each column vector $s$ to query the index.  
In addition to LSH, we also explore the more recent HNSW~\cite{DBLP:journals/pami/MalkovY20}.
HNSW is a proximity graph with multiple layers where two vertices are linked based on their proximity.
It supports fast nearest neighbor search with high recall.
We find that HNSW improves the query time by orders of magnitude and thus allows \name to support querying over the WDC corpus with 50M tables, which is much larger than the previously supported datasets for table union search.

Since such index structures return approximate instead of exact results, there might be some false negatives in the top-k results.
Nevertheless, we find in the experiments that the effectiveness loss caused by the false negatives is within a reasonable range.
Meanwhile, the query time can be reduced by one to three orders of magnitude (details in Section~\ref{subsec-scal}).

\subsection{Pruning Mechanism for Verification}\label{subsec-bound} 

%
Once a candidate table is found, we can reduce the expensive verification cost by quickly computing lower and upper bounds on the  unionability score.
We first look at how to estimate the upper bound $\textbf{UB}(S,T)$ between two tables $S$ and $T$.
Recall that in maximum weighted bipartite matching, each column/node in both $S$ and $T$ can be covered by at most 1 edge in the edges of the maximum matching.
If we remove this constraint, since nodes can appear in multiple edges, the new optimal matching is easy to compute. 
Moreover, as it allows edges with greater weights, the total score forms an upper bound of the true table unionability score $U(S, T)$. 
For the upper bound $\textbf{UB}(S,T)$, we first sort the edges by their weights in descending order. 
Then we add edges with the largest weights into the matching in a greedy manner. 
This process is repeated until all columns in $S$ or $T$ are covered or all edges are used.
The time complexity of the above process for calculating $\textbf{UB}(S,T)$ is $\bigO(|E| \log |E| + n)$, where $|E|$ is the number of edges in $G$.
It is much cheaper to compute than the real table unionability score.

Next, we introduce how to quickly estimate a meaningful lower bound $\textbf{LB}(S,T)$.
For lower bounds, we would like to find a set of edges that do not violate the constraint of bipartite matching, i.e., each column in the two tables is covered by one edge.
We can also achieve this goal via a greedy algorithm.
Similar to computing the upper bound, we sort the edges by weight in descending order and pick edges with the largest weights.
After that, we remove edges that are associated with the columns in the selected edges so as to avoid violations.
The termination condition of this process is also the same as that of calculating the upper bound.
Since the resulting matching does not necessarily cover all nodes in $S$ or $T$, the total weight $\textbf{LB}(S,T)$ is a lower bound of the maximum matching.
The time complexity of calculating $\textbf{LB}(S,T)$ is also $\bigO(|E| \log |E| + n)$.

\begin{example}
	We use the example in Figure~\ref{fig:bp} to illustrate the upper  bound computation.  \revision{Note this example is designed to illustrate the algorithm, not to model the actual distribution of weights in a data lake.}
	We fetch edges in the descending order of weight:
	$\langle s_1, t_2 \rangle$, $\langle s_1, t_1 \rangle$, $\langle s_2, t_2 \rangle$, and $\langle s_4, t_3 \rangle$. At this point, since all nodes $\{t_1, t_2, t_3\}$ in $T$ are covered, we stop here.
	The upper bound is $0.85+0.8+0.7+0.65 = 3$, larger than the exact value 2.15.
	
	To compute the lower bound, we start from edge $\langle s_1, t_2 \rangle$  and then remove all edges associated with $s_1$ and $t_2$.
	The remaining edge with maximum weight is $\langle s_4, t_3 \rangle$. 
	After involving this edge into the matching, there is no remaining one and the algorithm stops here.  
	Hence, the lower bound is $0.85+0.65 = 1.5$, which is smaller than the exact value 2.15. 
\end{example}

\section{Experiments}\label{sec:experiments}

We now present an evaluation of \name on real-world data lake corpora. 
First, we show that \name achieves new state-of-the-art results on table union search by outperforming the previous best methods by 6.8\% in MAP and Recall. 
Next, our scalability experiments show that \name (especially with the HNSW index) achieves significant performance gain (up to 3,000x) while preserving reasonable effectiveness performance.
Lastly, we conduct case studies to show how \name can generalize to another two dataset discovery applications: column clustering and table discovery for downstream machine learning tasks.
We include additional results and discussions in the appendix that is available in the full technical report~\cite{DBLP:journals/corr/abs-2210-01922}.

\subsection{Experiment Setup}\label{subsec-setup}

\subsubsection{Environment}
We implement \name in Python using Pytorch and the Hugging Face Transformers  library~\cite{DBLP:conf/emnlp/WolfDSCDMCRLFDS20}.
For contrastive learning, we use RoBERTa~\cite{DBLP:journals/corr/abs-1907-11692} as the base language model. 
We set the hyper-parameters batch size to 64, learning rate to 5e-5, and max sequence length to 256 across all the experiments.
All experiments are run on a server with configurations similar to those of a p4d.24xlarge AWS EC2 machine with 8 A100 GPUs.
The server has 2 AMD EPYC 7702 64-Core processors and 1TB RAM. 
\subsubsection{Datasets}\label{subsubsec:datasets}

We use five benchmark datasets with statistics detailed in Table~\ref{tab:bench_stats}. 
Firstly, we evaluate the effectiveness on the first three benchmark datasets, which are subsets of real Open Data.
\revision{Since accuracy requires manually labeled ground truth, such datasets are not very large. 
We only use them to conduct the experiments of effectiveness reported in Section~\ref{subsec-eff}.}
The \santosBench benchmark~\cite{santos23} consists of 550 real data lake tables drawn from 296 Canada, UK, US, and Australian open datasets, and 50 query tables. 
From Table Union Search~\cite{DBLP:journals/pvldb/NargesianZPM18}, there are two available benchmarks: 
\tusSmallBench and \tusLargeBench.
\tusSmallBench benchmark consists of 1,530 data lake tables that are derived from 10 base tables from Canada open data. 
We also use the larger benchmark, \tusLargeBench, which consists of $\sim$5,000 data lake tables derived from 32 base tables from Canada open data. 
For these two benchmarks, we randomly select 150 and 100 query tables, respectively, following previous studies~\cite{santos23,DBLP:journals/pvldb/NargesianZPM18}. 
The SANTOS\footnote{\url{https://github.com/northeastern-datalab/santos}} and \tus\footnote{\url{https://github.com/RJMillerLab/table-union-search-benchmark}} benchmarks, along with their ground truth of unionable tables, are publicly available.

The last two benchmark datasets are utilized in the experiments for efficiency and scalability.  
\revision{Compared with the previous three datasets, these two datasets do not have ground truth labels but have much larger cardinalities.}
The \realBench benchmark contains $\sim$11K raw data lake tables from Canada and UK open data, and 80 query tables. 
We also run experiments on the WDC web tables corpus~\cite{DBLP:conf/www/LehmbergRMB16} which contains 50.8 million relational web tables 
extracted from the Common Crawl. 
We randomly select 30 tables as the query.

\begin{table}[!ht]
\caption{\small Effectiveness (top) and scalability (bottom) benchmarks.}\vspace{-2em}
\label{tab:bench_stats}
\small
\begin{tabular}{lcccc}\\ \toprule
Benchmark & \# Tables & \# Cols  & Avg \# Rows   & Size (GB) \\ \midrule
\santosBench    & 550   & 6,322 & 6,921   & 0.45 \\
\tusSmallBench  & 1,530 & 14,810    & 4,466   & 1 \\
\tusLargeBench  & 5,043 & 54,923    & 1,915   & 1.5 \\ \midrule
\realBench    & 11,090   & 123,477    & 7,675   & 11 \\
WDC & 50M    & 250M    & 14  & 500 \\ \bottomrule
\end{tabular}\vspace{-2mm}
\end{table}

\subsubsection{Metrics}
For effectiveness, we perform evaluation based on the ground truth from the first three  benchmarks. 
For the \tus benchmarks, the tables are synthetically-partitioned from tables of distinct domains, so the ground truth is created in a generative manner.
As for the \santosBench benchmark, the tables have been manually-annotated to create a ground truth listing expected unionable tables to each query table.
Then we follow previous studies~\cite{santos23,DBLP:journals/pvldb/NargesianZPM18,DBLP:conf/icde/BogatuFP020,DBLP:books/daglib/0021593} and use the Mean Average Precision at k (MAP@k), Precision at k (P@k) and Recall at k (R@k) to evaluate the effectiveness in returning the top-k results. 
We compute each score by averaging 5 repeated runs.
For efficiency, we measure the average time per query. 

\subsubsection{Baselines}\label{subsubsec:baselines}
For effectiveness experiments, we compare our approach, \name, with the following existing approaches.

\noindent $\bullet$ \dtl~\cite{DBLP:conf/icde/BogatuFP020} extends Table Union Search~\cite{DBLP:journals/pvldb/NargesianZPM18} for the problem of finding related tables by using table features such as column names, value overlap, and formatting. 
	To compare fairly with \name, we omit the column name feature.
\\ $\bullet$ \santos~\cite{santos23} proposes an approach that leverages both columns and relationships between columns by using external and self-curated knowledge bases.
\\ $\bullet$ \slk~\cite{DBLP:conf/kdd/HulsebosHBZSKDH19} is a representation learning method that leverages several column features such as table statistics and word embeddings to learn the embedding vector of a column.
\\ $\bullet$ \sato~\cite{DBLP:journals/pvldb/ZhangSLHDT20} extends \slk by capturing the table context using LDA, and thus performing a form of multi-column prediction. \\
\revision{$\bullet$ \single 
is our column encoder proposed in Section~\ref{subsec:simclr} that only uses a single column as the input of the encoder in the training process.  This is \name without the use of contextual information from Section~\ref{subsec:contextualized}.}

For efficiency experiments, we aim at exploring the benefits brought by different design choices in the \name framework.
Thus we compare the performance of 4 methods: basic linear search (Linear), pruning based on estimated bounds (Pruning), search with an LSH index (LSH), and search with an HNSW index (HNSW).

\subsubsection{Column encoder settings}

We empirically choose the most suitable sampling method (Section~\ref{subsec:pretbl}) and augmentation operator 
(introduced in Section~\ref{subsec:contextualized} and more details in Appendix A). 
For sampling methods, we find that \name achieves the best performance when pre-trained with the cell-level TF-IDF scoring function on the \santosBench and \tusLargeBench benchmarks, and with a column-ordered sampling method, alphaHead, that sorts
tokens in alphabetical order performs the best, on \tusSmallBench.
For augmentation operators, we find that the drop\_col operator performs the best on \santosBench while
drop\_cell achieves the best performance on the two \tus benchmarks.

\subsection{Results for Effectiveness}\label{subsec-eff}

Table~\ref{tab:map} reports the results of MAP@k and R@k on the three benchmarks for all methods. 
Note that the results for \santos are unavailable for \tusLargeBench because \santos, which requires the labeled query table intent columns~\cite{santos23}, have not been evaluated on this benchmark due to the absence of annotated intent columns.
We run the experiments up to k=10 on \santosBench following~\cite{santos23}, and up to k=60 on the \tus benchmarks, which is consistent with~\cite{DBLP:journals/pvldb/NargesianZPM18}. 
\revision{Note the recall cannot reach 100\% when $k$ is smaller than the number of correct unionable tables from the labeled ground truth as reported in previous studies~\cite{DBLP:journals/pvldb/NargesianZPM18,santos23}.  For example, for \santosBench where $k$ is 10, the ground truth includes on average around 13 tables for different queries, so even the best technique can return (recall) at most 75\% or k of 10 of these.
Table~\ref{tab:map} indicates the maximum recall as IDEAL for each setting.}

\begin{table}[!ht]
\caption{\small MAP@k and R@k results on all benchmarks with ground truth, where k=10 for \santosBench benchmark and k=60 for the \tus benchmarks. \revision{The IDEAL R@k for \santosBench is 0.75, IDEAL R@k for \tusSmallBench is 0.341, and IDEAL R@k for \tusLargeBench is 0.277.}}\vspace{-1em}
\label{tab:map}
\small
\begin{tabular}{l|cc|cc|cc}\hline
         & \multicolumn{2}{c|}{\santosBench} & \multicolumn{2}{c|}{TUS                        Small} & \multicolumn{2}{c}{TUS Large} \\
\multicolumn{1}{c|}{Method} & MAP@k & R@k  & MAP@k & R@k & MAP@k & R@k \\  \hline \hline
\single     & 0.891   & 0.588 & 0.954 & 0.255 & 0.902 & 0.208\\
\sato          & 0.878   & 0.594 & 0.966 & 0.271 & 0.930 & 0.223\\
\slk      & 0.782   & 0.493 & 0.984 & 0.265 & 0.744 & 0.119\\
SANTOS        & 0.930   & 0.690 & 0.885 & 0.230 & -     & -\\
$D^3L$        & 0.523   & 0.422 & 0.794 & 0.215 & 0.484 & 0.124\\ 
Starmie       & \textbf{0.993}   & \textbf{0.737} & \textbf{0.991} & \textbf{0.277} & \textbf{0.965} & \textbf{0.238}\\ \hline
\end{tabular}
\end{table}

We can observe that \name outperforms the baselines across all three benchmarks. 
On the \santosBench benchmark, \name achieves the highest MAP@10 of 99.3\% and highest R@10 of 73.7\% (which is close to the IDEAL), outperforming \sato, \slk, \santos, \dtl baselines by large margins of 13\%, 27\%, 6.8\%, and 90\% respectively. 
Also, \name outperforms its \single variation by 11\%, showing that a multi-column approach is necessary. 
Similarly, on the \tusSmallBench benchmark, \name outperforms the highest-achieving baseline, \slk, by 0.7\% and \single variation by 4\% in MAP@k. 
On the \tusLargeBench benchmark, \name outperforms \sato by 4\% and \single by 7\% in MAP@k. 
Thus, the \name approach, by capturing column context and leveraging contrastive learning in pre-training, is very effective in solving the table union search problem.

\begin{figure}[t]
{
\centering
\begin{minipage}[t]{\textwidth}
\includegraphics[width=0.5\linewidth]{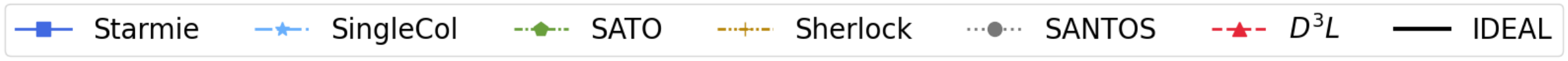}
\end{minipage}
}
\subfloat[$P@k$ on \santosBench]{
\begin{minipage}[t]{0.49\linewidth}
\includegraphics[width=\linewidth]{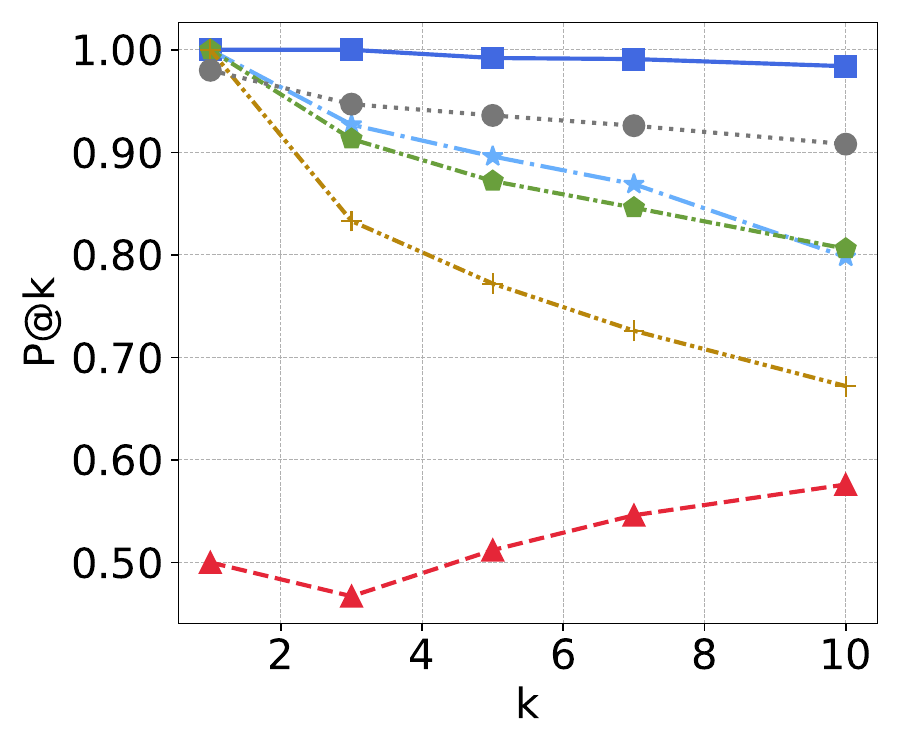}
\end{minipage}
}
\subfloat[$R@k$ on \santosBench]{
\begin{minipage}[t]{0.49\linewidth}
\includegraphics[width=\linewidth]{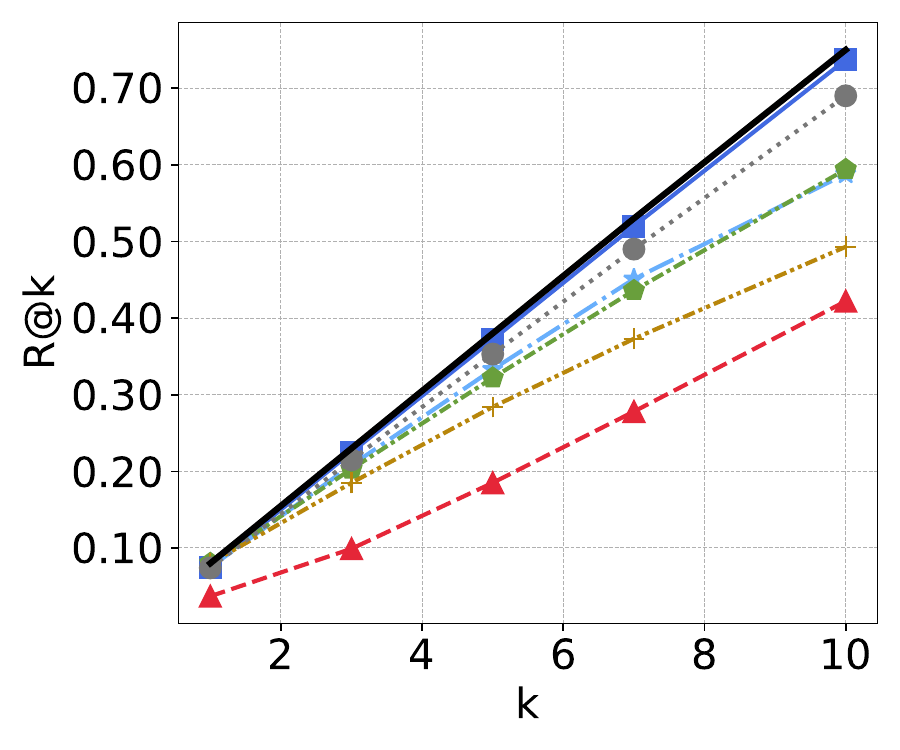}
\end{minipage}
}
\hfill
\subfloat[$P@k$ on TUS Small]{
\begin{minipage}[t]{0.49\linewidth}
\includegraphics[width=\linewidth]{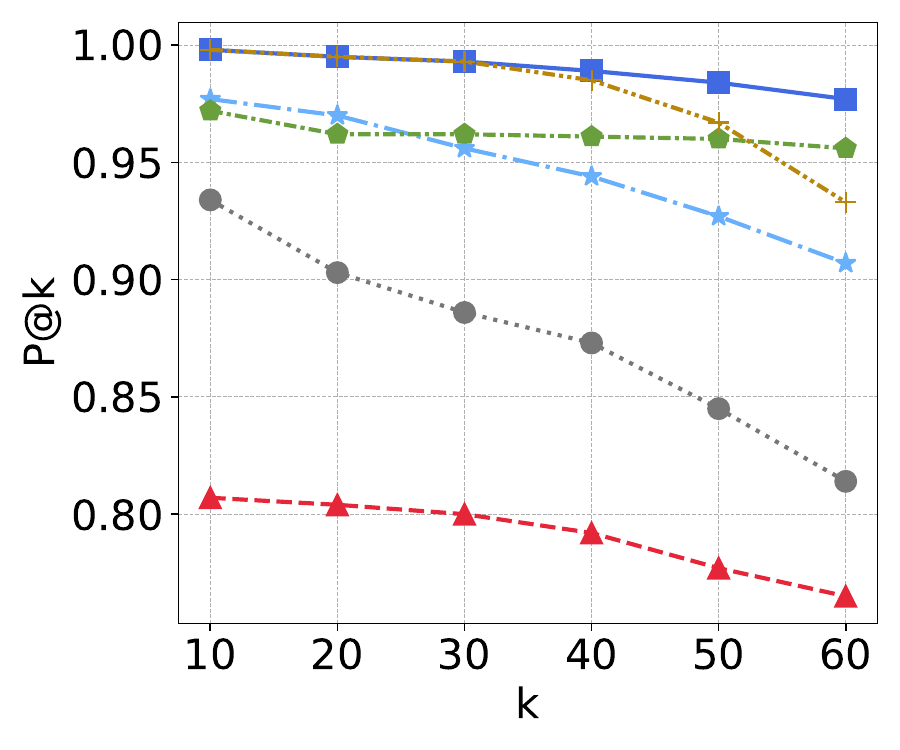}
\end{minipage}
}
\subfloat[$R@k$ on TUS Small]{
\begin{minipage}[t]{0.49\linewidth}
\includegraphics[width=\linewidth]{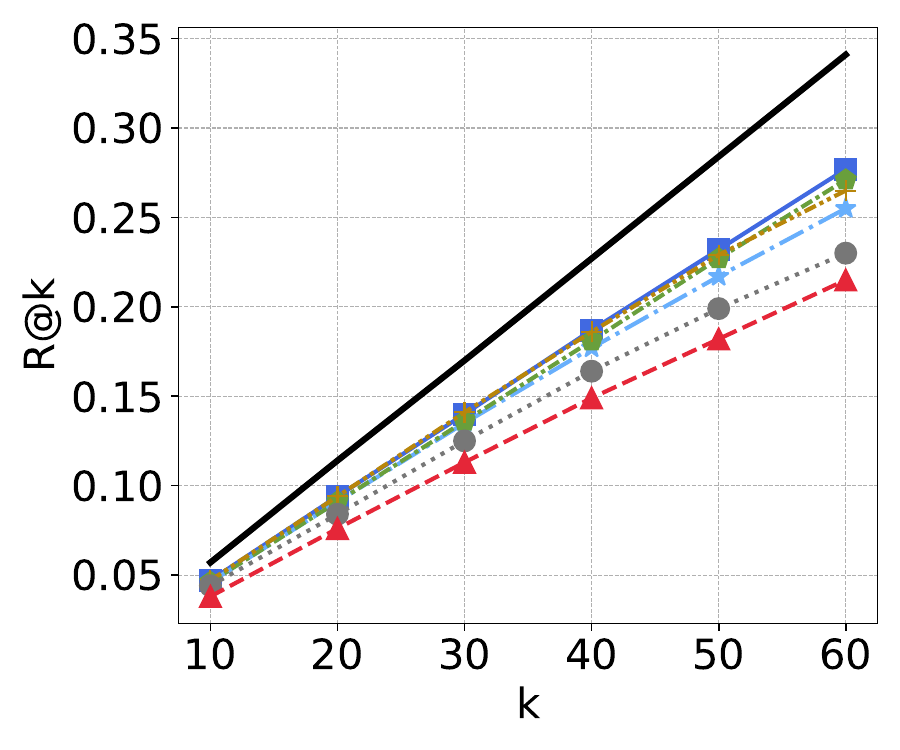}
\end{minipage}
}
\hfill
\subfloat[$P@k$ on TUS Large]{
\begin{minipage}[t]{0.49\linewidth}
\includegraphics[width=\linewidth]{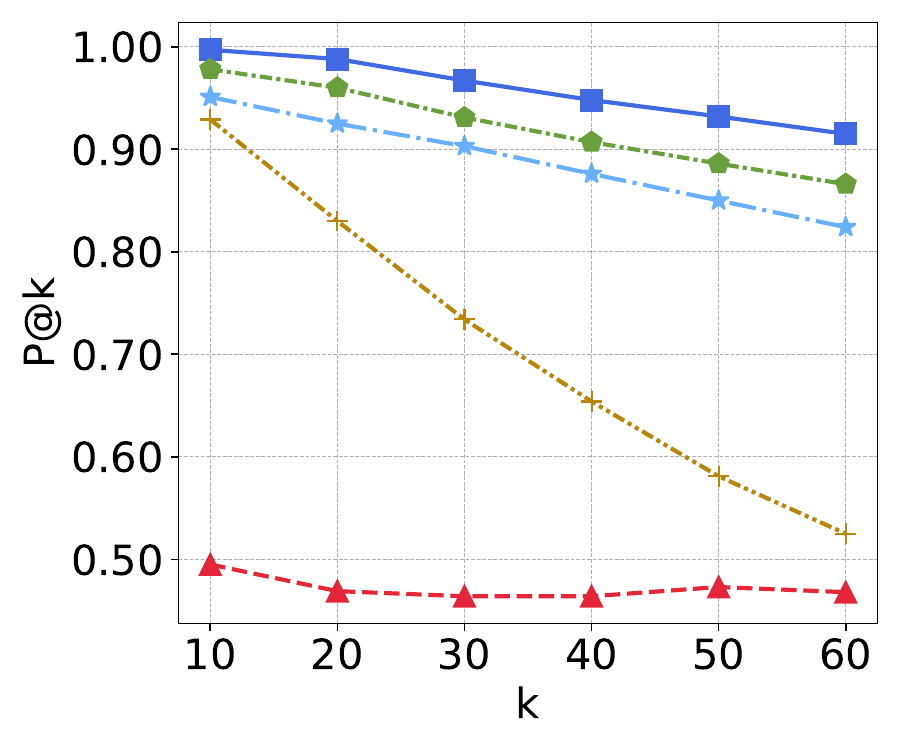}
\end{minipage}
}
\subfloat[$R@k$ on TUS Large]{
\begin{minipage}[t]{0.49\linewidth}
\includegraphics[width=\linewidth]{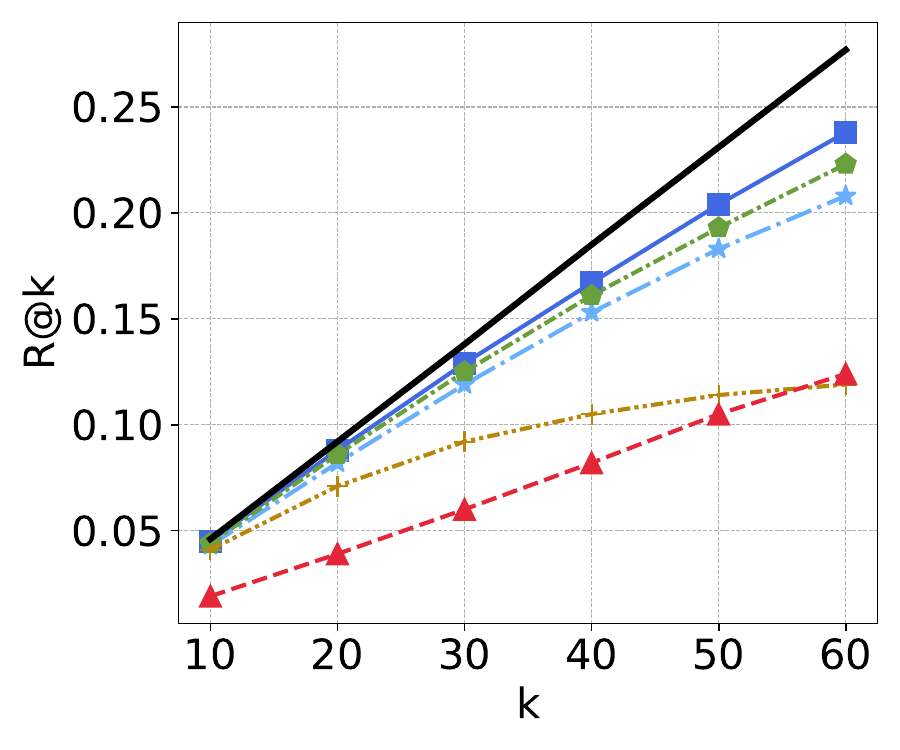}
\end{minipage}
}
\vspace{-1em}
\caption{$P@k$ and $R@k$ results on different benchmarks.}
\label{fig:p_r_all}
\end{figure}

Figure~\ref{fig:p_r_all} shows the P@k and R@k of \name and the baselines as k increases 
on all benchmarks. 
Throughout all values of k, \name outperforms all baselines for both P@k and R@k.
In Figures~\ref{fig:p_r_all}(b), (d), and (f), \name is closest to IDEAL, with R@10 only 1.8\% below IDEAL on \santosBench, R@60 18.8\% below IDEAL on \tusSmallBench, and R@60 14.1\% below IDEAL on \tusLargeBench.



\begin{figure*}[!ht]
    {
    \centering
    \begin{minipage}[t]{0.4\textwidth}
    \includegraphics[width=\linewidth]{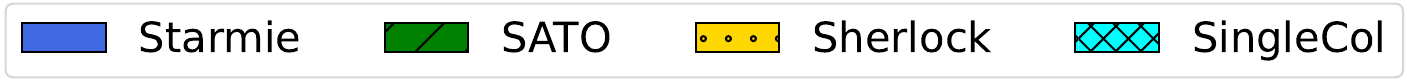}
    \end{minipage}
    }
    \subfloat[$MAP@k$ of different \# Cols]{
    \begin{minipage}[t]{0.3\linewidth}
    \includegraphics[width=\linewidth]{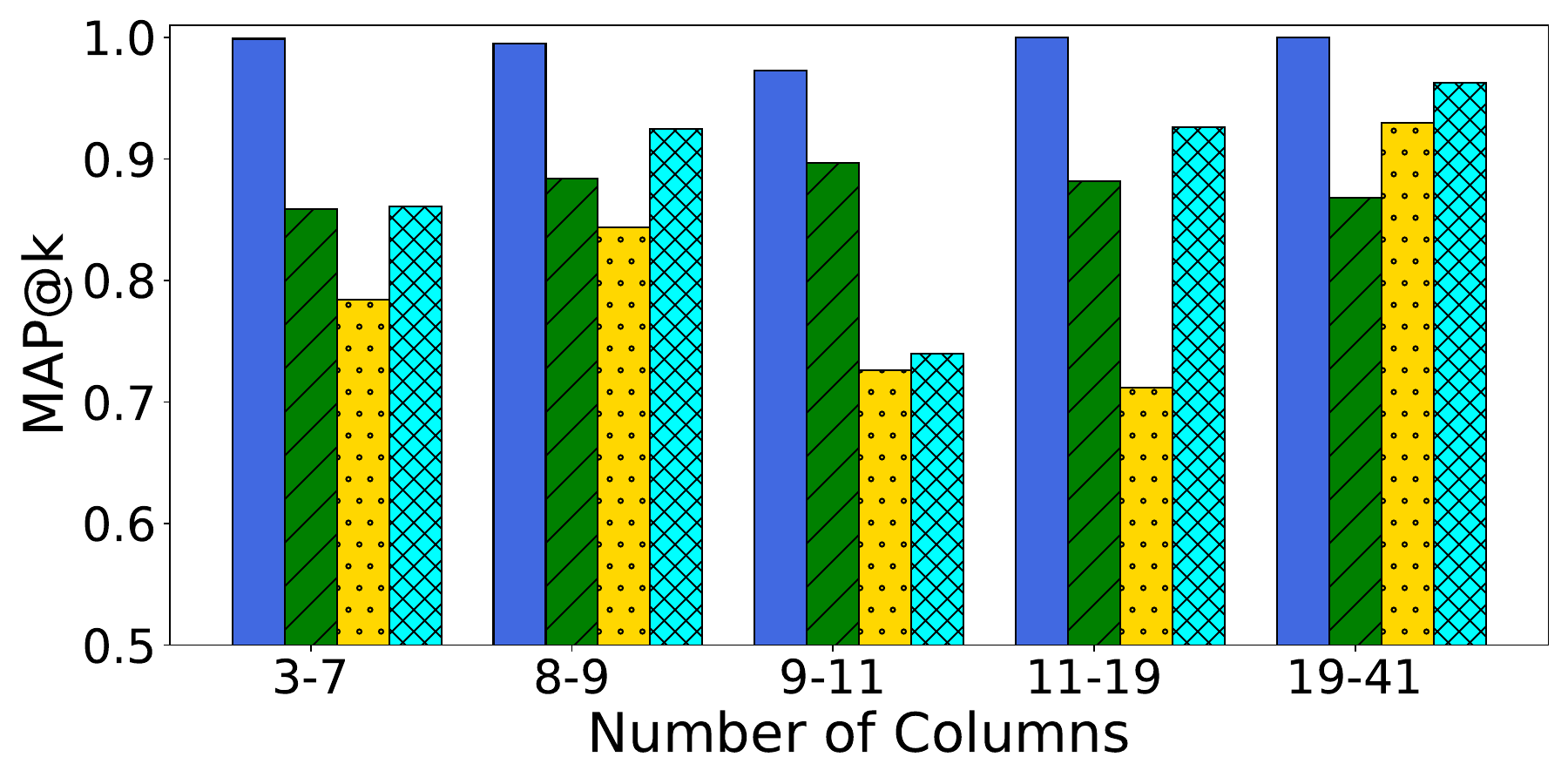}
    \end{minipage}
    }
    \subfloat[$MAP@k$ of different \# Rows]{
    \begin{minipage}[t]{0.3\linewidth}
    \includegraphics[width=\linewidth]{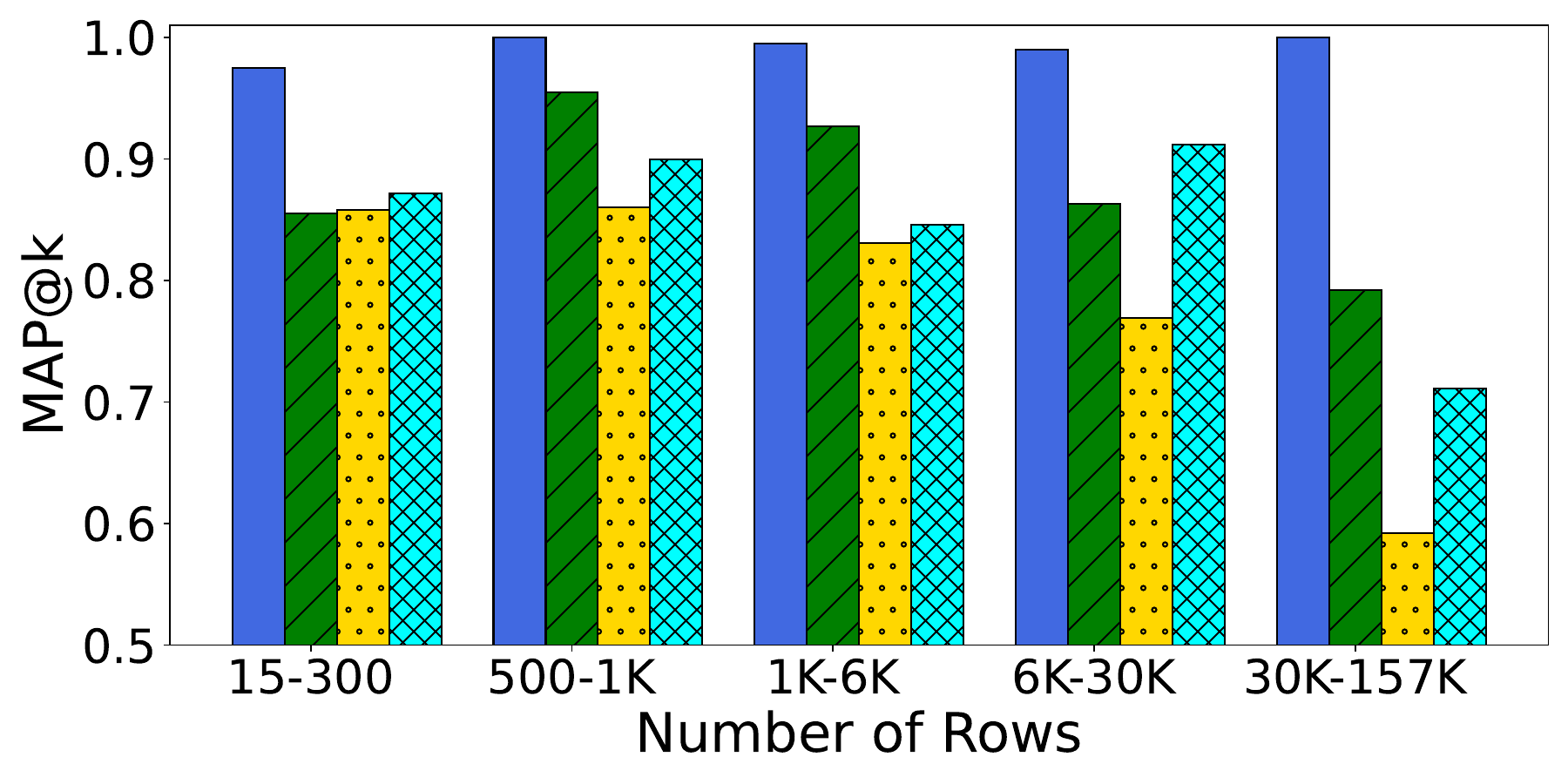}
    \end{minipage}
    }
    \subfloat[$MAP@k$ of different \% Num. Cols]{
    \begin{minipage}[t]{0.3\linewidth}
    \includegraphics[width=\linewidth]{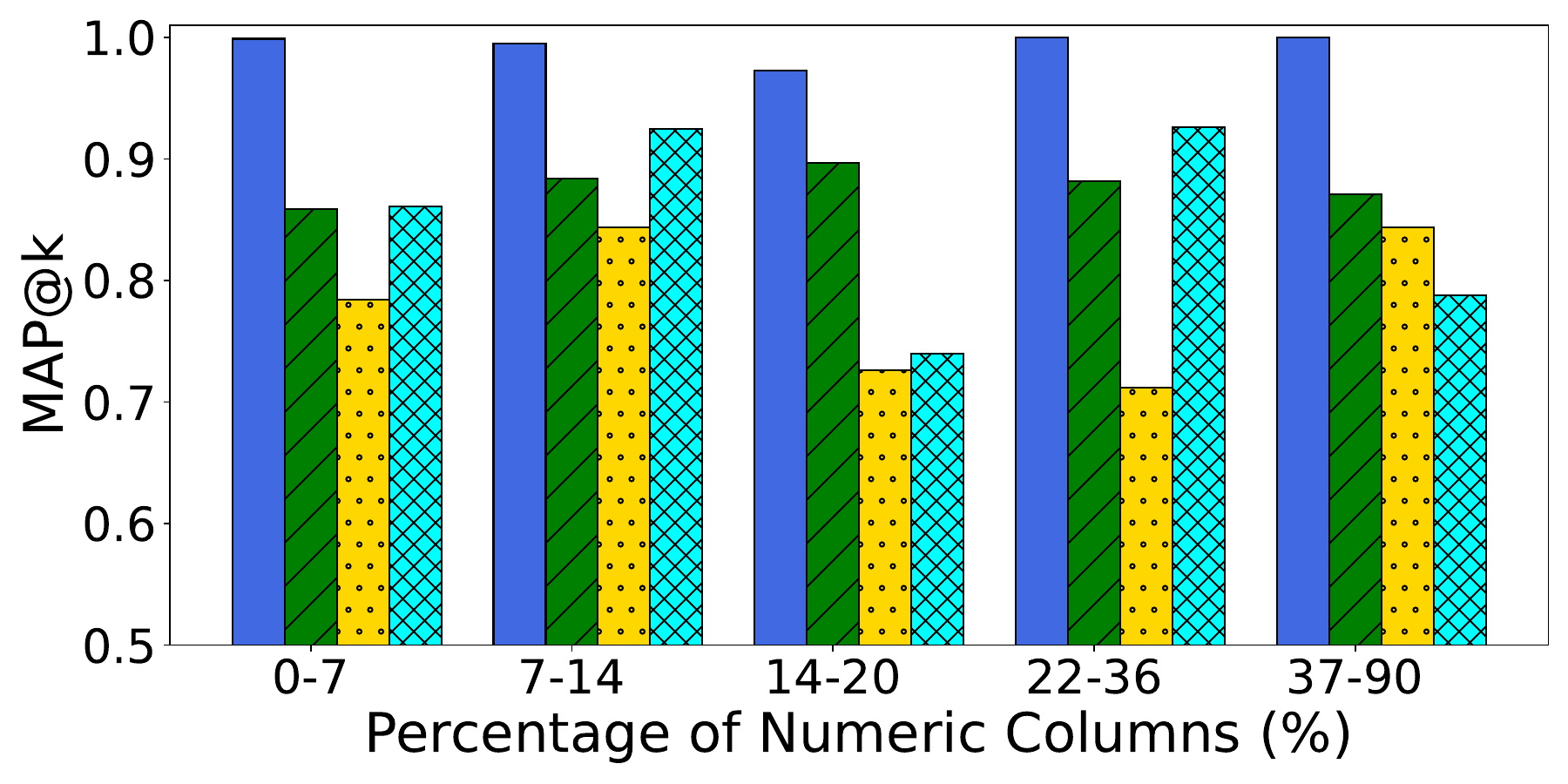}
    \end{minipage}
    }
    \caption{In-depth analysis of \name, \sato, \slk, \revision{and \single} as we vary the number of columns, number of rows, and percentage of numerical columns on the \santosBench benchmark.}
    \vspace{-2mm}
    \label{fig:error_santos}
\end{figure*}

To better understand the influence of datasets on the performance of \name, we conducted an in-depth analysis to look at its performance for different settings of arity, cardinality, and percentage of numerical columns in query tables. 
We evenly split the query tables into five groups \revision{for each setting}.
\revision{We compare \name with alternative representation methods \sato, \slk, and \single  that also use deep learning to encode columns into high-dimensional vectors.}
\revision{As shown in Figure~\ref{fig:error_santos}(a)/(c), \name consistently outperforms the baselines as the number of columns are varied and as the percentage of numeric columns varies.   
As the number of rows increases (Figure~\ref{fig:error_santos}(b)), the results of \name remain consistently high while the performances of \sato, \slk, \revision{and \single} generally decrease. 
We believe this is due to our efforts table preprocessing techniques (Section~\ref{subsec:pretbl}). }
\revision{Meanwhile, the performance of \single is much worse than \name under all settings, which illustrates the importance of contextual information in training the column encoders.}
The methods have similar trends on the \tusSmallBench and \tusLargeBench (Appendix B).

\revision{
\subsubsection{Micro Benchmarking Experiment}\label{subsubsec:micro}

To evaluate the effect of self-supervision, specifically randomly drawing two tables to create negative examples on the effectiveness of \name, we create a microbenchmark consisting of eight data lakes drawn from \tusSmallBench benchmark.  
In each data lake, there are 470 tables, of which 25\% of tables have the same class as the query table while the remaining 75\% of tables are evenly divided among 2-9 negative classes of tables. As shown in Table~\ref{tab:microbenchmark}, for data lakes with fewer classes, it is less likely that two random tables are not unionable (counterexamples), thus resulting in lower MAP scores compared to those with more classes. Still, the MAP remains high, showing that effect of assuming random tables are not unionable is negligible -- even in the extreme case of a data lake with only 3 classes of tables.  
We report MAP for K of 60, which is consistent with other experiments on \tusSmallBench benchmark, and K of 120, which shows a clearer trend of MAP increasing and stabilizing as the number of classes increases. Note that this benchmark contains up to only 10 class labels, so the fluctuating trend is only among a limited set of classes.  Real lakes would contain many magnitudes more classes.
}
\begin{table}[ht]
	\caption{Effectiveness of \name on data lakes with different numbers of classes of tables.}\vspace{-1em}
    \label{tab:microbenchmark}
    \begin{tabular}{ccccccccc}\toprule
        & \multicolumn{8}{c}{\# of Negative Classes} \\
         & 2 & 3 & 4 & 5 & 6 & 7 & 8 & 9 \\ \midrule
    MAP@60 & 0.99 & 1.0 & 1.0 & 1.0 & 1.0 & 1.0 & 1.0 & 1.0 \\
    MAP@120 & 0.89 & 0.93 & 0.94 & 0.95 & 0.93 & 0.94 & 0.92 & 0.92 \\\bottomrule
    \end{tabular}
\end{table}

\subsection{Scalability}\label{subsec-scal}

\begin{table}[!ht]
\caption{Effectiveness of different design choices. The first four methods are for \name. }\vspace{-2em}
\small
\label{tab:metrics_runtime}
\begin{tabular}{lcccc}\\ \toprule
\multicolumn{1}{c}{Method} & MAP@10 & P@10 & R@10 & Query Time (s) \\ \midrule
Linear      & 0.993 & 0.984 & 0.737 & 96\\
Pruning      & 0.993 & 0.984 & 0.737 & 61\\
LSH Index   & 0.932 & 0.780 & 0.580 & 12\\
HNSW Index  & 0.945 & 0.810 & 0.606 & 4\\
\midrule
\sato        & 0.878 & 0.806 & 0.594 & 252\\ 
\slk    & 0.782 & 0.672 & 0.493 & 264\\
\single
 & 0.891 & 0.798 & 0.588 & 108\\ \bottomrule
\end{tabular}
\end{table}

\smallskip\noindent
\textbf{Impacts on effectiveness. }
Since some design choices might result in  effectiveness loss, we report their results of three evaluation metrics on the \santosBench benchmark. 
As shown in Table~\ref{tab:metrics_runtime}, we compare \name with a basic linear scan with  three other design choices (above the horizontal line), as well as baselines \sato, \slk, and \single (full experiment results are shown in Appendix C). 
The main takeaway is that HSNW  preserves the effectiveness as much if not better than the LSH index that is widely used in previous studies, while having tremendous speed improvement. 
This suggests HSNW is a very promising direction for providing real-time search over massive data lakes.

\noindent
\textbf{Preprocessing time. }
Since \name requires model pre-training and model inference, in addition to possibly indexing, we provide some insights of such overhead by comparing its preprocessing time with existing systems \dtl and \santos that are not based on pre-trained LMs. 
The preprocessing time of \name consists of the following parts: pre-training taking 3.1 hours, model inference taking 4.4 min, and indexing taking 10-30 sec. 
Meanwhile, \dtl takes 7.6 hours to create four indexes for each column feature and \santos takes 17 hours to create indexes using a knowledge base and the data lake.
Thus, pre-training a language model in \name does not incur too much overhead compared to existing systems.

\begin{figure*}[!ht]
    {
    \noindent\hspace{0.25\linewidth}
    \begin{minipage}[t]{\textwidth}
    {\includegraphics[width=0.5\linewidth]{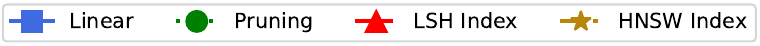}}
    \end{minipage}
    }
    \subfloat[\realBench benchmark]{
    \begin{minipage}[t]{0.49\linewidth}
    \includegraphics[width=\linewidth]{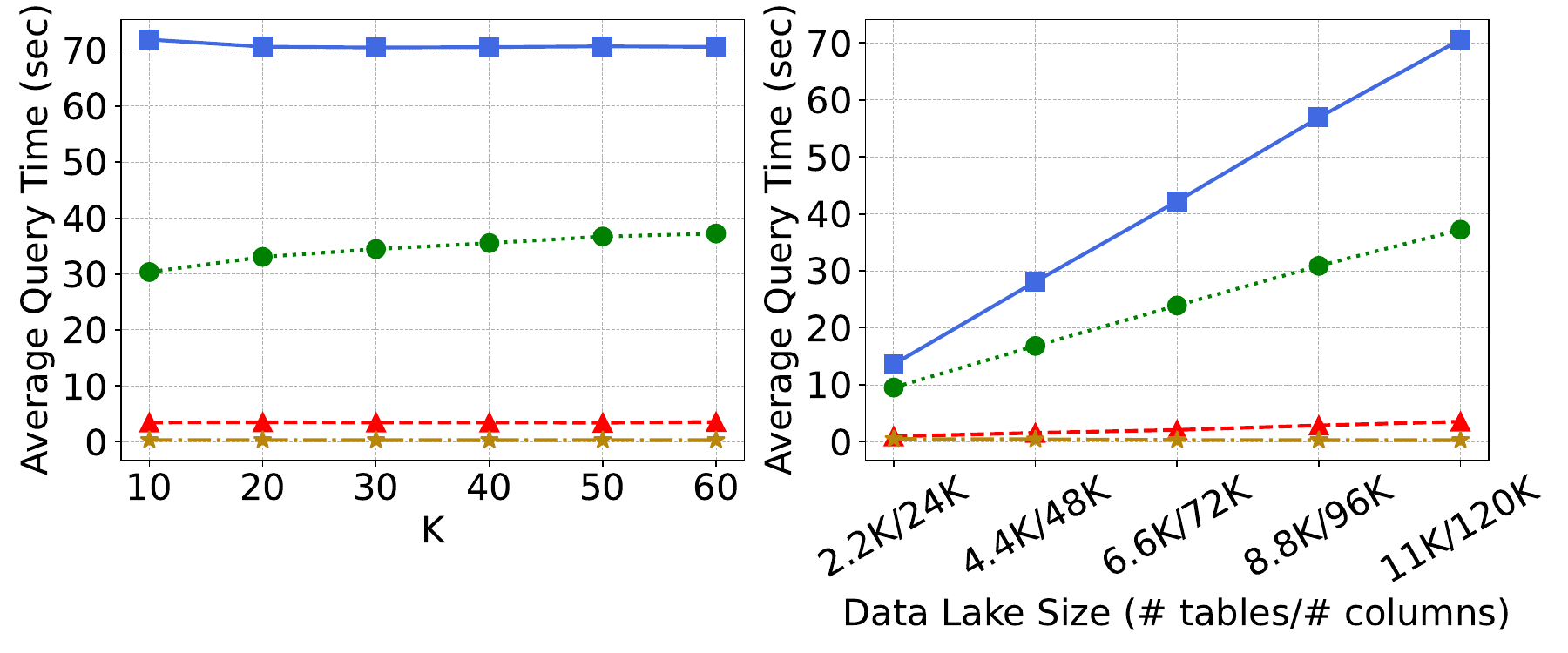}
    \end{minipage}
    }
    \subfloat[WDC (Sample) benchmark]{
    \begin{minipage}[t]{0.25\linewidth}
    \includegraphics[width=\linewidth]{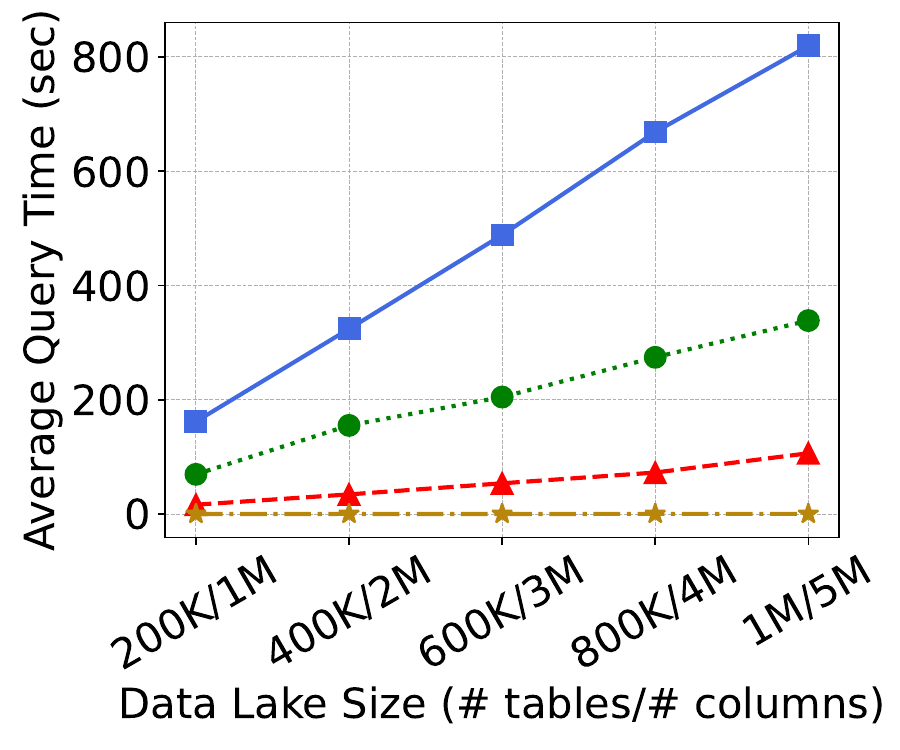}
    \end{minipage}
    }
    \subfloat[WDC (Full) benchmark]{
    \begin{minipage}[t]{0.25\linewidth}
    \includegraphics[width=\linewidth]{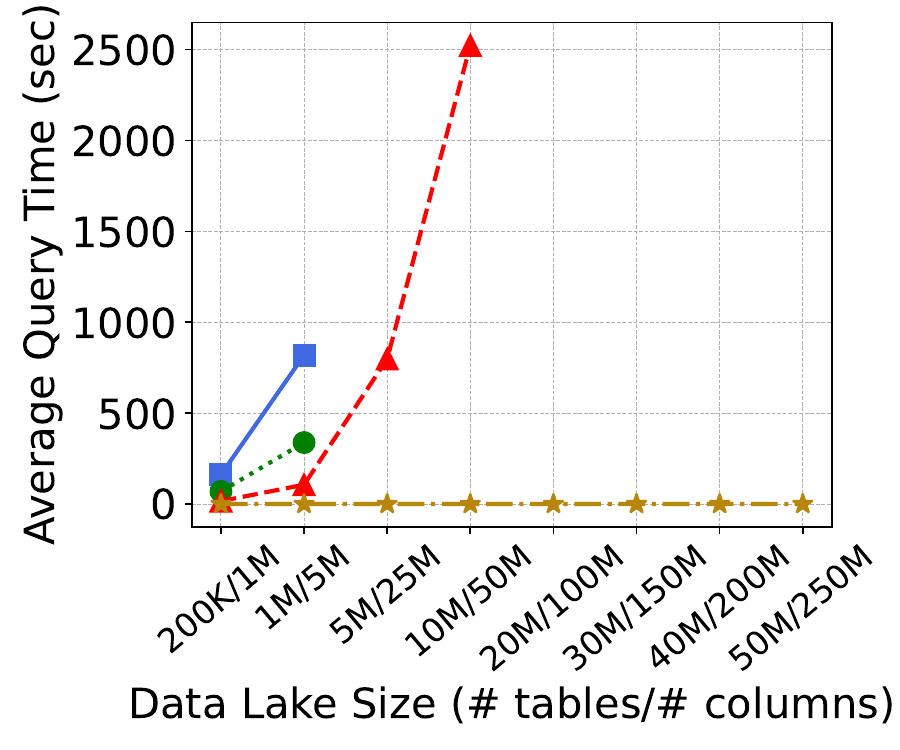}
    \end{minipage}
    }
    \vspace{-1em}
    \caption{Scalability on the \realBench benchmark, a sample of 1M WDC tables, and the full WDC benchmark}
    \label{fig:scal}
\end{figure*}

\noindent
\textbf{Time efficiency. }
We have observed that the employed design choices can speed up the online query time while sufficiently preserving the effectiveness scores. 
Next we evaluate the scalability of different design choices. 
In Figure~\ref{fig:scal}(a), we first evaluate the four variations of \name on the \realBench benchmark, as we increase the number of returned unionable tables k from 10 to 60. 
We then evaluate their query times as the data lake size grows to its full size of $\sim$11K tables / $\sim$120K columns. 
We also experiment on the WDC benchmark, specifically when the data lake grows to 1M tables / 5M columns (Figure~\ref{fig:scal}(b)) to show the trend of each method , and when the data lake grows to 50M tables / 250M columns (Figure~\ref{fig:scal}(c)). 
For each method, if a data point's query time does not finish within 24 hours, then we consider it as timeout and omit the result from the corresponding figures.
\revision{To show the effectiveness of the LB/UB pruning mechanisms proposed in Section~\ref{subsec-bound}, we compare the number of verification steps needed 
per query with and without pruning.
We found that on the SANTOS Small benchmark, the average number of verifications for Linear (without LB/UB pruning) is 550,  while that of Pruning is 342 (38\% reduction). 
This result shows that the pruning heuristic indeed helps significantly reduce the unnecessary verification and thus improves the overall performance.}

Throughout all these experiments, we see that the design choice with the HNSW index leads to the best performance. 
On the \realBench benchmark in Figure \ref{fig:scal}(a), the k-scalability experiment shows that Pruning is 2X faster than Linear, while LSH index is 20X faster than Linear. 
Meanwhile, HNSW index, which leads to an average query time of around 300 ms, is 220X faster than Linear and 11X faster than the popular LSH index. 
As the data lake grows to its full size, there is a steady increase in query time of Linear and Pruning; while that of LSH index and HNSW index remain stable, with the query time of HNSW index remaining around 400 ms. 
On the WDC benchmark in Figure \ref{fig:scal}(b), there is a similar trend as the data lake grows to 1M tables. 
On the full WDC benchmark in Figure \ref{fig:scal}(c), Linear and Pruning time out after 1M tables, while LSH index times out after having an average query time of 2,520 sec on 10M tables. 
Meanwhile, the query time for HNSW index stays consistent at around 60 ms as the data lake grows to its full size of 50M tables / 250M columns.
The reason is that the hierarchical graph-based structure of HNSW allows it to locate to the nearest neighbors much faster than hash-based indexes~\cite{DBLP:journals/pami/MalkovY20}.
Overall, the design choices explored in this paper, especially HNSW index, show a great improvement in the average query time, even when the data lake grows to an immense size of 50M tables.
Meanwhile, to the best of our knowledge, the largest dataset that are evaluated by existing solutions of table union search is with only 5,000 tables / 1M columns~\cite{DBLP:journals/pvldb/NargesianZPM18}, which has 250 times smaller number of columns.

\noindent
\textbf{Memory overhead. }
Lastly, we examine the relative memory overhead of \name with different design choices (No index denotes linear scan and pruning methods from Table~\ref{tab:metrics_runtime}).
In Table~\ref{tab:storage}, we report the memory usage of \name relative to the total data lake size (11 GB) of
\realBench. 
The results show that \name is not only scalable but also memory efficient:  its variations take up around 3-7\% space overhead.
The memory saving is mainly due to the condensed vector column representations of \name which take up only 3\% of the original data lake size.

\begin{table}[ht]
	\caption{Relative Memory Overhead on the \realBench benchmark for \name with the data lake of 11 GB. }\vspace{-1em}
	\label{tab:storage}
	\begin{tabular}{lcc}\toprule
		Method & Memory Usage & Space Overhead \\ \midrule
		No Index      & 359 MB    & 3.26\% \\
		LSH Index   & 733 MB    & 6.66\% \\
		HNSW Index  & 749 MB    & 6.81\% \\\bottomrule
	\end{tabular}
\end{table}\vspace{-1em}

\subsection{Data discovery for ML tasks}

Next, we conduct a case study to show that \name can be applied to another application scenario of dataset discovery, i.e., retrieving relevant tables to improve
the performance of downstream ML tasks. 
For this case study, we consider a subset of 78k WDC tables
used in the evaluation of \sato~\cite{DBLP:journals/pvldb/ZhangSLHDT20}, from which
we collect all the 4,130 tables of at least 50 
rows as the data lake tables. Among these tables, we find that 25 tables of at least 200 rows
contain a numeric column called ``Rating''. These 25 tables contain various types of ratings
including those for sportsmen, TV shows, US congress members, etc.
From these tables, we construct 25 regression tasks with the goal of training an ML model that predicts
``Rating'' as the target column.
Since the ratings are from different domains, we normalize their values to the range $[0, 1]$.
More details about the setting can be found in Appendix D.

\begin{figure}[!ht]
    \centering
    \includegraphics[width=0.48\textwidth]{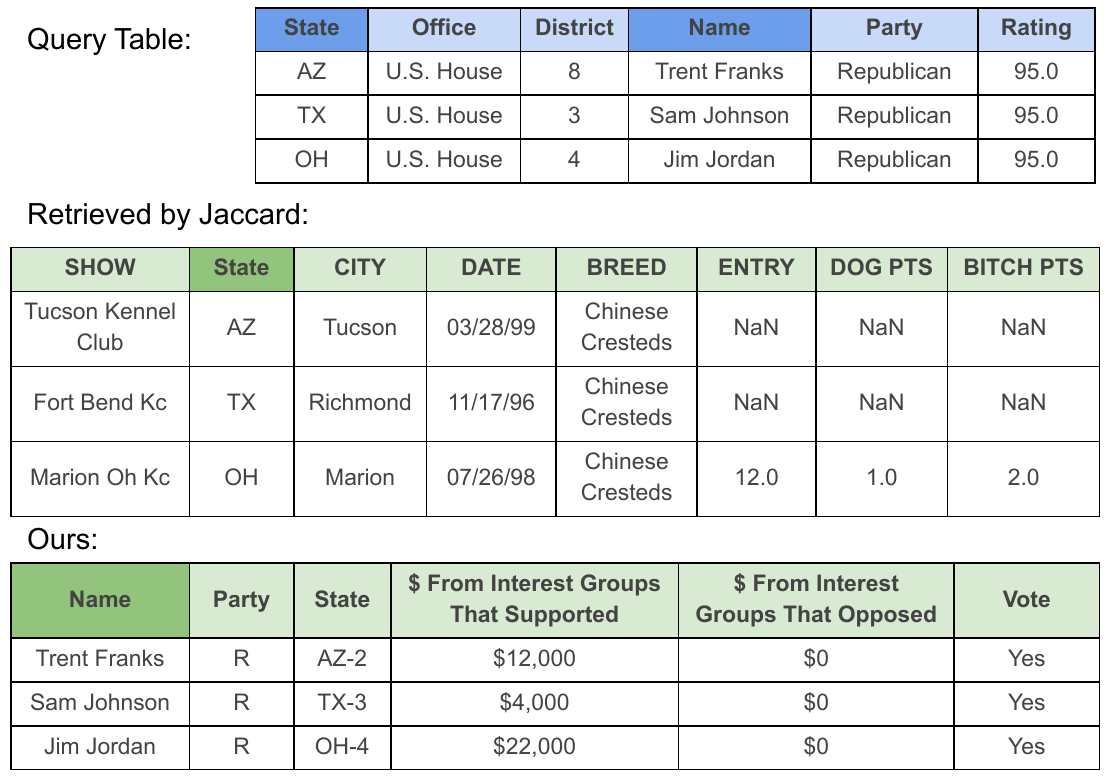}
    \caption{\small Example tables retrieved by Jaccard vs. \name.
    By joining the query table with the DL table retrieved by \name,
    the MSE for predicting the ``Rating'' attribute drops from 0.1598
    to 0.1195 (vs. 0.1544 when joining with the table retrieved by Jaccard).
    }
    \label{fig:ml_tables}\vspace{-3mm}
\end{figure}

For each task, we train a Gradient-Boosted Tree model~\cite{DBLP:conf/kdd/ChenG16} with all non-target
columns as features. We featurize the textual columns using Sentence Transformers~\cite{DBLP:conf/emnlp/ReimersG19}. We split each dataset into training and test sets at a ratio of 4:1. Note that the original dataset  may not contain informative features. 
Figure \ref{fig:ml_tables} shows such a dataset of US congress members. 

To improve the model's performance on these downstream tasks, we leverage \name to retrieve relevant tables from the data lake to join with the datasets (i.e., the query tables) to provide additional features. 
To showcase the effectiveness of \name, we use \name's contextualized column embeddings to retrieve from the data lake table that contains a column having the highest cosine similarity with a non-target column of the query table.
Finally, we augment the query table by performing 
a left-join with the retrieved table to ensure that the size of the augmented table stays unchanged.
We also consider two popular similarity methods for this task, Jaccard and Overlap~\cite{DBLP:conf/sigmod/ZhuDNM19,DBLP:conf/icde/DongT0O21}, as baselines
by replacing the cosine similarity scores with the corresponding similarity functions.

Table \ref{tab:ml} summarizes the results of the 3 evaluated methods. While all 3 methods result in
performance improvement (i.e., reduction of MSE), \name achieves significantly better
overall improvements with a 14.75\% MSE reduction, on 15/25 tasks improved, and by an average of 20.64\%. 
By inspecting the retrieved tables, we find that \name indeed retrieves qualitatively better candidate
tables. As Figure \ref{fig:ml_tables} shows, for the same US congress members table,
Jaccard similarity \revision{retrieves} an irrelevant table of dog competitions that also contains
a similar ``State'' column, but the two tables are not semantically relevant.
On the other hand, \name retrieves a table consisting of the amount of money raised from different
interest groups, which is a potentially relevant feature to ``Rating''. Indeed, by joining
with the retrieved table by \name, the MSE of the model drops from 0.1598 to 0.1198 (by $>$25\%).

\begin{table}[!hb]
\vspace{-0.5em}
	\caption{\small Performance gain of data discovery methods on 25 rating prediction tasks from WDC.}\vspace{-1em}
	\label{tab:ml}
	\begin{tabular}{ccccc}\toprule
		& NoJoin & Jaccard & Overlap & \name      \\ \midrule
		Avg. MSE     & 0.0820 & 0.0753  & 0.0748  & 0.0699  \\
		Improvement  & -      & 8.23\%  & 8.82\%  & \textbf{14.75\%} \\
		\#improved   & -      & 13      & 12      & \textbf{15}      \\
		avg. Improve & -      & 14.74\% & 14.05\% & \textbf{20.64\%} \\ \bottomrule
	\end{tabular}\vspace{-4mm}
\end{table}

\subsection{Case study: Column clustering}

Finally we show another application scenario of \name in dataset discovery: column clustering.
Specifically, we apply \name as a column encoder to provide embeddings 
for clustering all the 119,360 columns from the 78k WDC tables 
used in the experiments of \slk, \sato, and others~\cite{DBLP:conf/kdd/HulsebosHBZSKDH19,DBLP:journals/pvldb/ZhangSLHDT20,DBLP:conf/sigmod/SuharaL0ZDCT22}. These columns are annotated with 78 ground truth semantic types
such as population, city, name, etc.
The goal of column clustering is to discover clusters 
of columns that are semantically relevant.
The task of semantic type detection has traditionally been solved as 
a supervised multi-class classification problem
which requires significant annotated training data~\cite{DBLP:conf/sigmod/SuharaL0ZDCT22}. 
\name provides an \emph{unsupervised} solution.
From the contextualized column embeddings, we can  construct a similarity graph over all data lake columns as nodes. 
We can then add undirected edges between all pairs of columns having cosine similarities above a 
threshold $\theta$ (e.g., 0.6). 
Next the column clusters can be generated via any graph clustering algorithm. We choose the connected component algorithm for efficiency and simplicity. 

With \name, the clustering algorithm generates 2,297 clusters with an average cluster size of 51.96.
We measure the quality of the clusters by the \emph{purity} score, which is the percentage of columns assigned with the same semantic type as the majority ground truth
type of each cluster. 
The discovered clusters are generally of high quality as they achieve a purity score of 51.19 while using baselines such as \slk and \sato only achieves 30.5 or 37.36 purity scores when generating a similar number of clusters.
A more detailed example of discovered clusters is shown in Appendix E.

\section{Related Work}\label{sec:related}

\subsection{Dataset Discovery}

Dataset Discovery has been a hot topic in the data management community.
Earlier studies~\cite{DBLP:journals/pvldb/CafarellaHWWZ08,DBLP:journals/pvldb/VenetisHMPSWMW11,DBLP:journals/pvldb/AdelfioS13} relied on keyword search over web tables to identify essential information.
\textsf{Octopus}~\cite{DBLP:journals/pvldb/CafarellaHK09} and \textsf{InfoGather}~\cite{DBLP:conf/sigmod/YakoutGCC12} focused on the problem of schema complement, an important topic in exploring web tables.
\textsf{Aurum}~\cite{DBLP:conf/icde/FernandezAKYMS18}, \textsf{S3D}~\cite{DBLP:conf/cikm/GalhotraK20} and \textsf{Tableminer+}~\cite{DBLP:conf/semweb/MazumdarZ16,DBLP:journals/semweb/Zhang17} utilized knowledge bases to identify relationship between datasets.
\textsf{SemProp}~\cite{DBLP:conf/icde/FernandezMQEIMO18} followed this route by leveraging ontologies and word embeddings, and \textsf{Leva}~\cite{DBLP:conf/sigmod/ZhaoF22} solved a similar problem with graph neural networks.
\textsf{$D^4$}~\cite{DBLP:journals/pvldb/OtaMFS20} addressed the problem of column clustering in data lake tables.
\textsf{Valentine}~\cite{DBLP:conf/icde/KoutrasSIPBFLBK21} provided resources for evaluating column matching tasks.
\textsf{DomainNet}~\cite{DBLP:conf/edbt/LeventidisRMRG21} studied the problem of disambiguation in data lakes.

Finding related tables from data lakes is an essential task in dataset discovery.
There are two sub-tasks in this application, namely finding joinable tables and table union search~\cite{DBLP:conf/sigmod/SarmaFGHLWXY12}.
To support finding joinable tables, earlier studies utilized syntactic similarity metrics that are widely used in the applications of string similarity search and join~\cite{DBLP:conf/icde/WuZWLFX19,DBLP:conf/icde/HarmouchPN21,DBLP:conf/icde/LiLL08}.
\textsf{LSH Ensemble} used containment (overlap)~\cite{DBLP:journals/pvldb/ZhuNPM16} as the similarity metric and provided a high-dimensional similarity search based solution.
\textsf{Josie}~\cite{DBLP:conf/sigmod/ZhuDNM19} employed overlap over tokens and developed an exact data-optimized solution.
\textsf{PEXESO}~\cite{DBLP:conf/icde/DongT0O21} relied on cosine similarity over word embeddings and proposed indexing techniques to improve performance.
The table union search problem has been well explored recently.
Ling et al.~\cite{DBLP:conf/ijcai/LingH0Y13} and Lehmberg et al.~\cite{DBLP:journals/pvldb/LehmbergB17} illustrated the importance of finding unionable Web tables.
Nargesian et al.~\cite{DBLP:journals/pvldb/NargesianZPM18} proposed the first definition and comprehensive solution for the table union search problem in data lakes.
Bogatu et al.~\cite{DBLP:conf/icde/BogatuFP020} proposed the \textsf{$D^3$L} system by dividing columns into different categories.
The \santos~\cite{santos23} system uses a knowledge base along with binary relationships in the data lake to identify tables that share unionable columns and relationships, and it is the state-of-the-art approach in this field.
To the best of our knowledge, our work is the first solution to utilize contrastive learning techniques in table union search.

\subsection{Representation Learning for Tables}

Recently  many efforts  use representation learning techniques to address  problems related to tabular data.
\textsf{Sherlock}~\cite{DBLP:conf/kdd/HulsebosHBZSKDH19} and \textsf{Sato}~\cite{DBLP:journals/pvldb/ZhangSLHDT20} used a supervised feature based approach to learn vector representations for tables and columns.
\textsf{TURL}~\cite{DBLP:journals/pvldb/DengSL0020} proposed to use a pre-trained language model for web table related tasks and to come up with benchmark datasets for several tasks.
And pre-trained language models have been widely applied to different table-related applications, including entity matching~\cite{DBLP:journals/pvldb/0001LSDT20,DBLP:conf/sigmod/CappuzzoPT20,DBLP:journals/jdiq/LiLSWHT21}, column type detection~\cite{DBLP:conf/www/WangSLHDJ21,DBLP:conf/sigmod/SuharaL0ZDCT22}, and question answering~\cite{DBLP:conf/acl/YinNYR20,DBLP:conf/naacl/IidaTMI21}.
Our work follows this line of study and proposes the first solution that employs a pre-trained language model in a fully unsupervised way for the problem of table union search.

\section{Conclusion and Future Work}\label{sec:conclusion}

In this paper, we mainly focused on the problem of table union search, an essential application in dataset discovery from data lakes.
We argued that it is crucial to utilize contextual information to determine whether two columns are unionable and proposed \name, an end-to-end framework based on contrastive representation learning as the solution.
We also developed a multi-column table Transformer encoder that can capture the contextual information from a table so as to learn contextualized column embeddings.
Experimental results on popular benchmark datasets demonstrated that \name significantly outperformed existing solutions for table union search.

\revision{
Our results show the promise of self-supervised contrastive learning in improving the accuracy of table union search, as well as joinable table search, and column clustering -- the latter areas we are exploring further.  
We believe the improved accuracy justifies the use of learning over previous heuristic approaches and the self-supervision will be important to data lakes where labeled training data is expensive to collect and generalize.  Our results using the relatively new HNSW index are exciting and important in the development of real-time data lake search solutions.  
}


\balance
\bibliographystyle{ACM-Reference-Format}
\bibliography{ref/core,ref/table,ref/other}

 \newpage

\appendix 
\section{Optimizing Table Preprocessing}\label{app:operators}


As such, the default table preprocessing method can fail to capture the most relevant information from the input table for the downstream tasks. 
To address this issue, in \name, we explore a design space for tuning and optimizing this process to make hyper-parameter tuning and future exploration easier.

To start, we first determine whether to read the table horizontally (row by row) or vertically (column by column).
Since table union search typically relies on column alignment, we assume that
the column-ordered method will achieve better performance, which is verified empirically in Section \ref{sec:experiments}.
We assume the column-order approach for the rest of the design options.
Next we explore a reasonable solution from the following three aspects.

\smallskip
\noindent\textbf{Token/cell scoring functions. } First of all we need to decide how to score the importance of each token or cell for each column.
Here we consider the TF-IDF method for token scoring, where the importance of each token is computed as its inverse document frequency $\log(M) / |\{t \ | \ \mathsf{token} \in t\}|$, where $t$ is a column and $M$ is the number of all data lake columns. 
Then the cell score is obtained by summing or averaging the TF-IDF scores of tokens in it.

\smallskip
\noindent\textbf{Deterministic vs. non-deterministic. } After obtaining the score of each cell, we sort cells in the descending order of importance scores for each column. 
Next, we can select and concatenate the tokens/cells either in a deterministic manner, e.g., in the descending order of the importance score, until we reach the token budget for each column (the max length uniformly distributed among columns), or a non-deterministic one by sampling the tokens/cells with probability proportional to their importance scores. 
We try both ways in our experiments.

\smallskip
\noindent\textbf{Row alignment. } Finally, we need to align the selected cells in a column.
If we simply concatenate the top-ranked tokens or cells, the row-alignment information from the original table might be lost after preprocessing.
In other words, the order of cells in the serialized columns may not follow the same order of the rows. 
Row alignment information can be  useful, e.g., (``\texttt{California}'', ``\texttt{Sacramento}'') and (``\texttt{New York}'', ``\texttt{Albany}'') as in our example for capturing the state-capital relation.
We propose another option that ranks all the rows by their average cell-level scores and then selects the rows to be included in the serialization result in either deterministic or non-deterministic ways.

\section{Comparing different operators}\label{sec-ablation}
\subsection{Augmentation Operators}\label{subsec-augAblation}
\begin{figure}[h!t]
    \centering
    \includegraphics[width=0.48\textwidth]{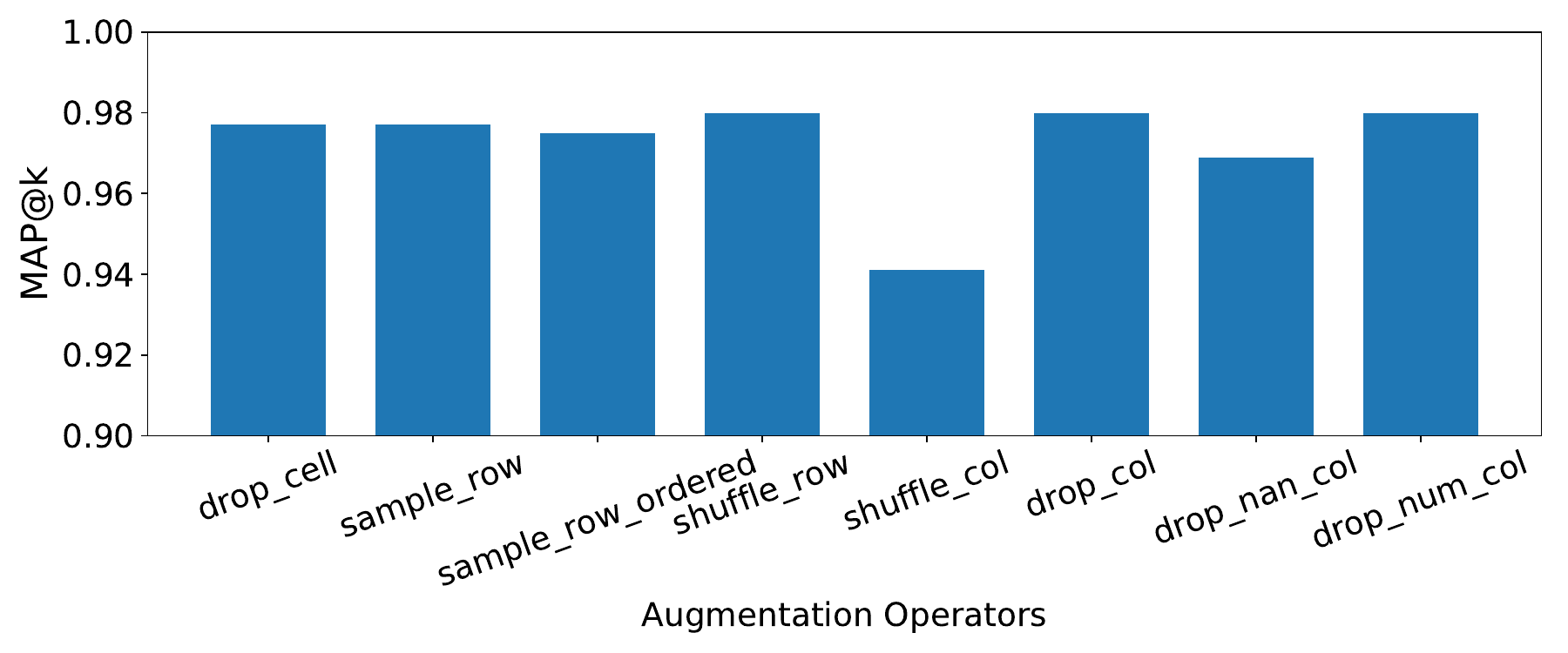}
    \caption{MAP@10 results on \santosBench benchmark using different augmentation operators.}
    \label{fig:ablation_ao}
\end{figure}

To find the most effective augmentation operator used in pre-training (Section~\ref{subsec:contextualized}) on the \santosBench benchmark, we conduct experiments 
comparing the MAP@k scores of different $\mathsf{op}$'s, shown in Figure 
\ref{fig:ablation_ao}. 
Specifically, we experiment with augmentation operators at different table levels, including some of the operators listed in Table \ref{tab:da}:

\smallskip\noindent
\textbf{Cell-Level: }
\begin{itemize}
    \item drop\_cell: drops a random cell in a column
\end{itemize}

\smallskip\noindent
\textbf{Row-Level: }
\begin{itemize}
    \item sample\_row: samples a random percentage of the rows
    \item sample\_row\_ordered: samples random percentage of the rows, while preserving the original order of the rows
    \item shuffle\_row: shuffles the row order
\end{itemize}

\smallskip\noindent
\textbf{Column-Level: }
\begin{itemize}
    \item shuffle\_col: shuffles the column order
    \item drop\_col: drops a random subset of column
    \item drop\_nan\_col: drops columns consisting mostly of NaN's
    \item drop\_num\_col: drops a random subset of numeric columns
\end{itemize}

From this ablation study, we find that the column-level operator drop\_col leads to the highest MAP@k of 98\%, and thus conduct the effectiveness experiments with the drop\_col $\mathsf{op}$.

\subsection{Sampling Methods}
\begin{figure}[!ht]
    \centering
    \includegraphics[width=0.48\textwidth]{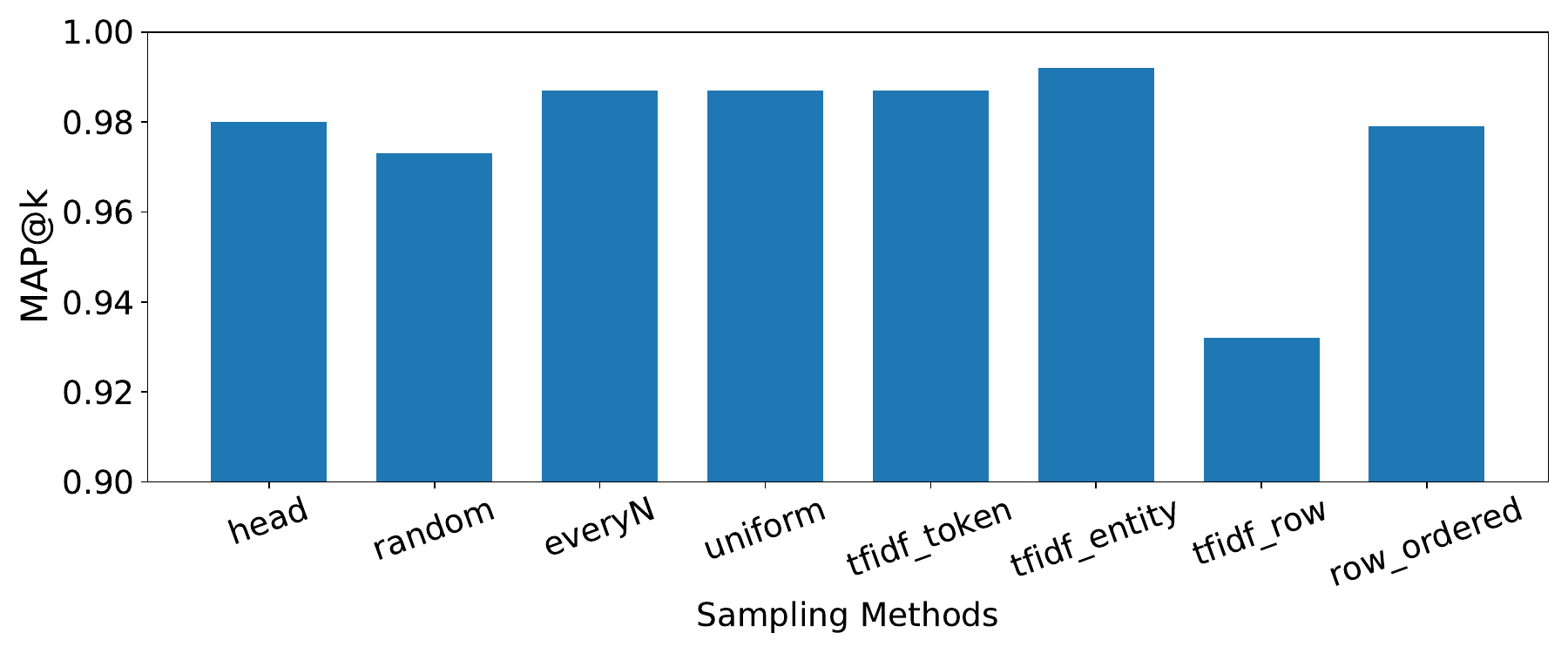}
    \caption{MAP@10 results on \santosBench benchmark using different sampling methods. }
    \label{fig:ablation_sm}
\end{figure}

We also conduct an empirical study comparing different sampling methods to find the method that best preserves the most meaningful tokens in table preprocessing. Specifically, we experiment with the following sampling methods, categorized by the level of the table. Note that all methods preserve the original order of the tokens/cells/rows, while taking unique samples:


\smallskip\noindent
\textbf{Column-Based, Token-Level: }
\begin{itemize}
    \item head: sample first N tokens
    \item random: randomly sample tokens
    \item everyN: sample every Nth token 
    \item uniform: sample most frequently-occurring tokens
    \item tfidf\_token: sample tokens with highest TF-IDF
    \item alphaHead: sample first N tokens sorted alphabetically scores
\end{itemize}

\smallskip\noindent
\textbf{Column-Based, Cell-Level: }
\begin{itemize}
    \item tfidf\_entity: sample cells in a column with highest average TF-IDF scores over its tokens
\end{itemize}

\smallskip\noindent
\textbf{Row-Level: }
\begin{itemize}
    \item tfidf\_row: samples rows with highest average TF-IDF scores over tokens in a row
    \item row\_ordered: sample and serialize tokens in a row
\end{itemize}

For the design space listed in Section \ref{app:operators}, we reach the following conclusions from the results shown in Figure \ref{fig:ablation_sm}:

\smallskip\noindent
\textbf{Row-ordered vs. column-ordered: } Out of all the sampling methods, the only row-ordered method is ``row\_ordered" (tfidf\_row is column-ordered but selects cells based on the highest average TF-IDF score across the row). The column-ordered methods outperform row\_ordered, with the highest column-ordered method tfidf\_entity achieving a MAP@k of 99.3\% while row\_ordered has a MAP@k of 97.9\%, thus confirming the original hypothesis.

\smallskip\noindent
\textbf{Token/cell scoring functions: } So far we have experimented with simple scoring functions (e.g. head, random), with the most complex scoring function being TF-IDF. However, we can see that the TF-IDF-based methods, specifically tfidf\_entity performs the best.

\smallskip\noindent
\textbf{Deterministic vs. non-deterministic: } All methods except for ``random" are deterministic. Since the best-performing deterministic method, tfidf\_entity, outperforms the non-deterministic method ``random" (which achieves a MAP@k of 97.3\%) we conclude that deterministic methods are more effective.

\smallskip\noindent
\textbf{Row alignment: } Methods tfidf\_row and row-ordered preserve the row alignment. We can see that column alignment is still more effective, but this design space requires further experimentation.

All in all, this ablation study on the \santosBench benchmark shows that the sampling method tfidf\_entity performs the best, with a MAP@k of 99.3\%. Thus, we conduct our effectiveness experiments on the \santosBench benchmark with tfidf\_entity as the sampling method.

\section{In-depth Analysis on Effectiveness}\label{sec-error_analysis}
\begin{figure*}[!ht]
    {
    \centering
    \begin{minipage}[t]{0.3\textwidth}
    \includegraphics[width=\linewidth]{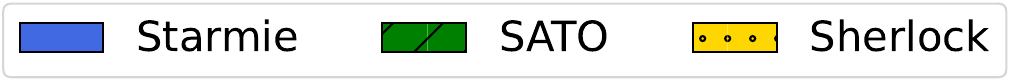}
    \end{minipage}
    }
    \subfloat[$MAP@k$ of different \# Cols]{
    \begin{minipage}[t]{0.3\linewidth}
    \includegraphics[width=\linewidth]{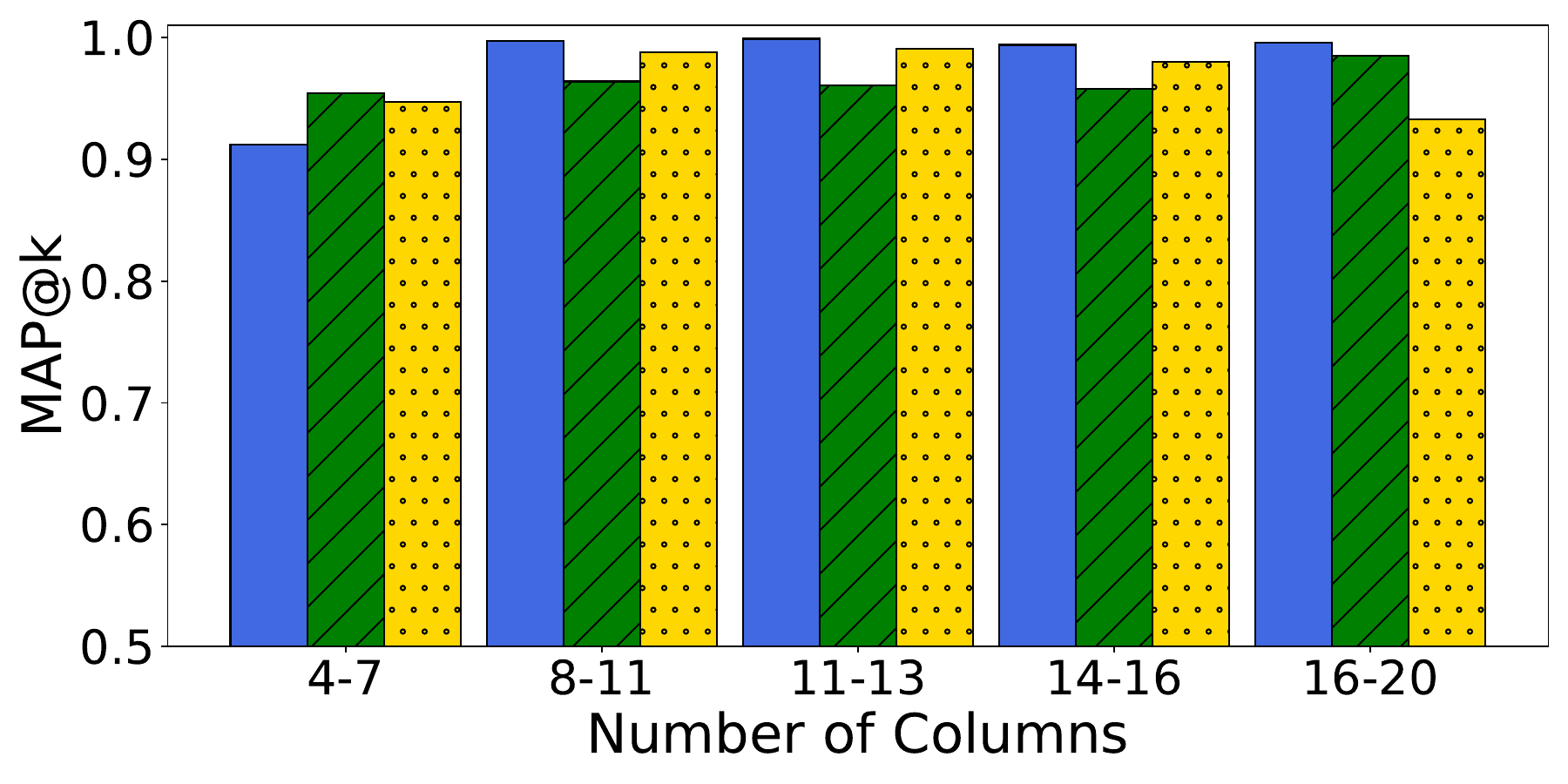}
    \end{minipage}
    }
    \subfloat[$MAP@k$ of different \# Rows]{
    \begin{minipage}[t]{0.3\linewidth}
    \includegraphics[width=\linewidth]{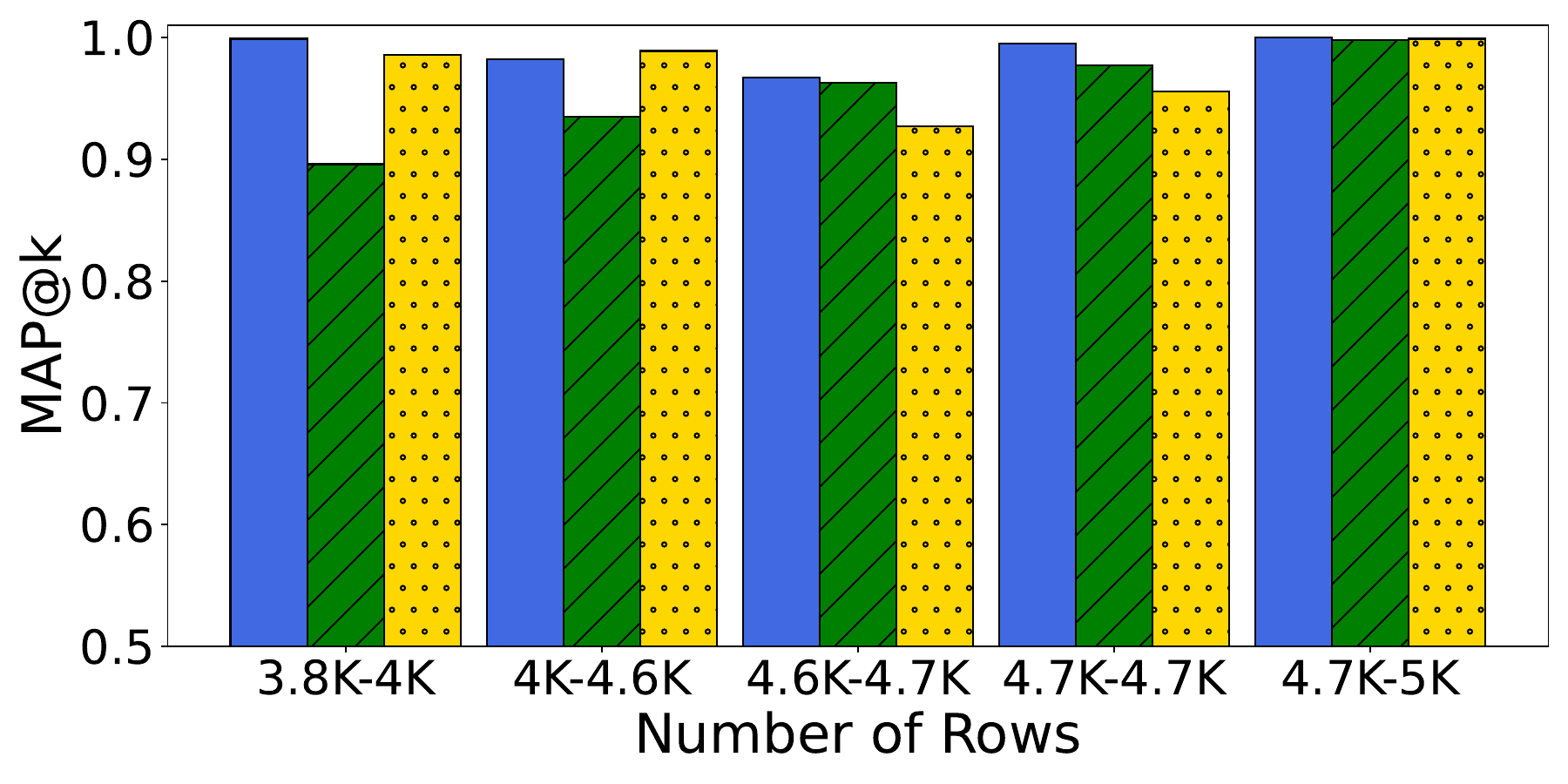}
    \end{minipage}
    }
    \subfloat[$MAP@k$ of different \% Num. Cols]{
    \begin{minipage}[t]{0.3\linewidth}
    \includegraphics[width=\linewidth]{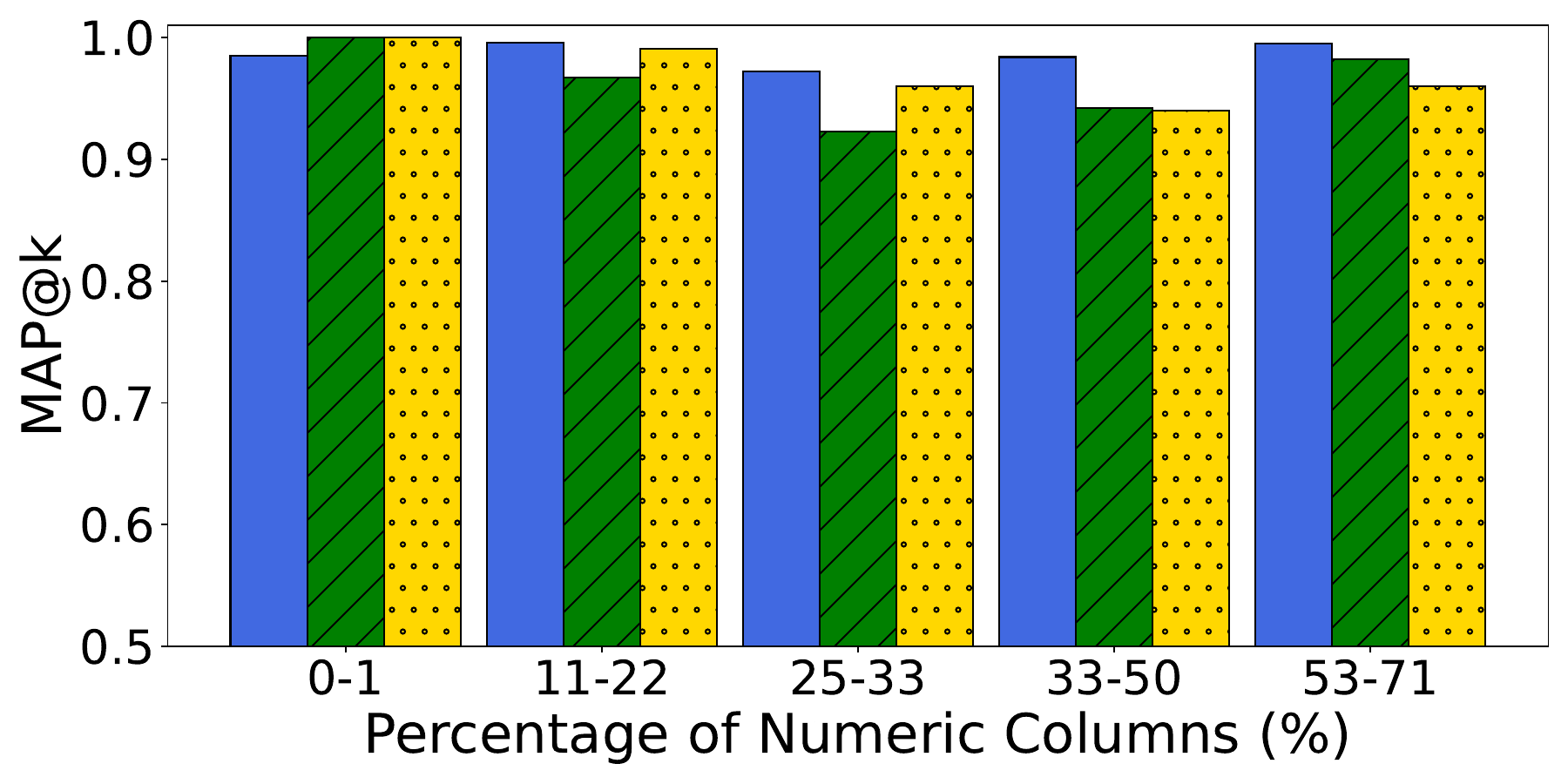}
    \end{minipage}
    }
\caption{In-depth Analyses of \name, SATO, and Sherlock as we vary the number of columns, number of rows, and percentage of numerical columns on the \tusSmallBench benchmark.}
\label{fig:error_tus_small}
\end{figure*}

Similar to the in-depth analyses conducted for \santosBench benchmark, shown in Figure \ref{fig:error_santos}, we conduct experiments on both TUS benchmarks to analyze \name's performance compared to the baselines SATO and Sherlock. With the same three analyses per benchmark, and the same division of tables into 5 buckets, we first explore the MAP@k results on the \tusSmallBench benchmark. In Figure \ref{fig:error_tus_small}, \name again outperforms the baselines and is robust to data containing large numbers of columns, rows, and high percentage of numeric columns. The baselines also show relatively consistent results across all buckets in the three analyses, which can be attributed to the fact that the \tusSmallBench benchmark is derived from only 10 seed tables, and thus may not exhibit much heterogeneity. This hypothesis requires further analysis.

\begin{figure*}[!ht]
    {
    \centering
    \begin{minipage}[t]{0.3\textwidth}
    \includegraphics[width=\linewidth]{fig/error_analyses/analyses_legends_no_ideal.pdf}
    \end{minipage}
    }
    \subfloat[$MAP@k$ of different \# Cols]{
    \begin{minipage}[t]{0.3\linewidth}
    \includegraphics[width=\linewidth]{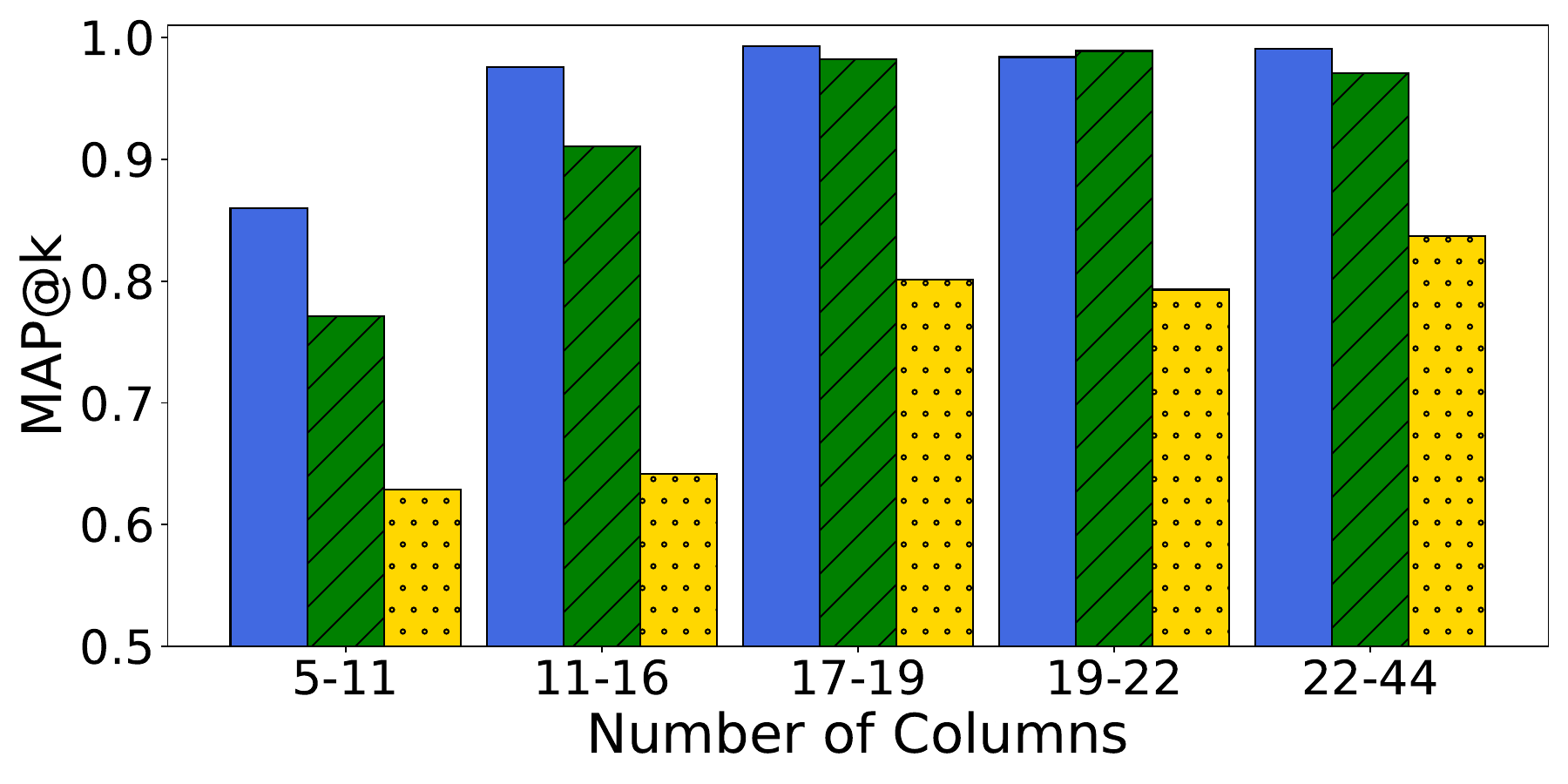}
    \end{minipage}
    }
    \subfloat[$MAP@k$ of different \# Rows]{
    \begin{minipage}[t]{0.3\linewidth}
    \includegraphics[width=\linewidth]{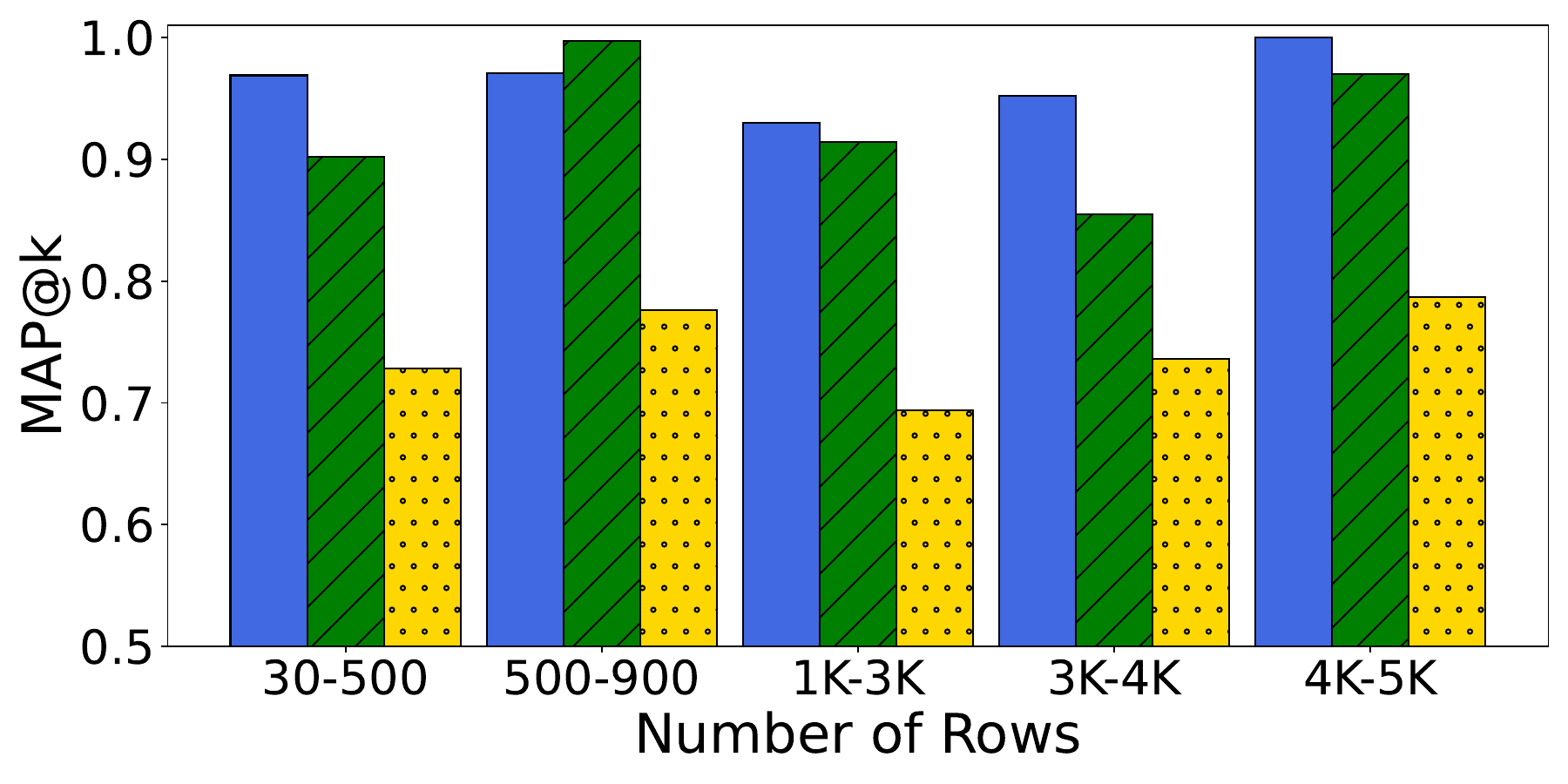}
    \end{minipage}
    }
    \subfloat[$MAP@k$ of different \% Num. Cols]{
    \begin{minipage}[t]{0.3\linewidth}
    \includegraphics[width=\linewidth]{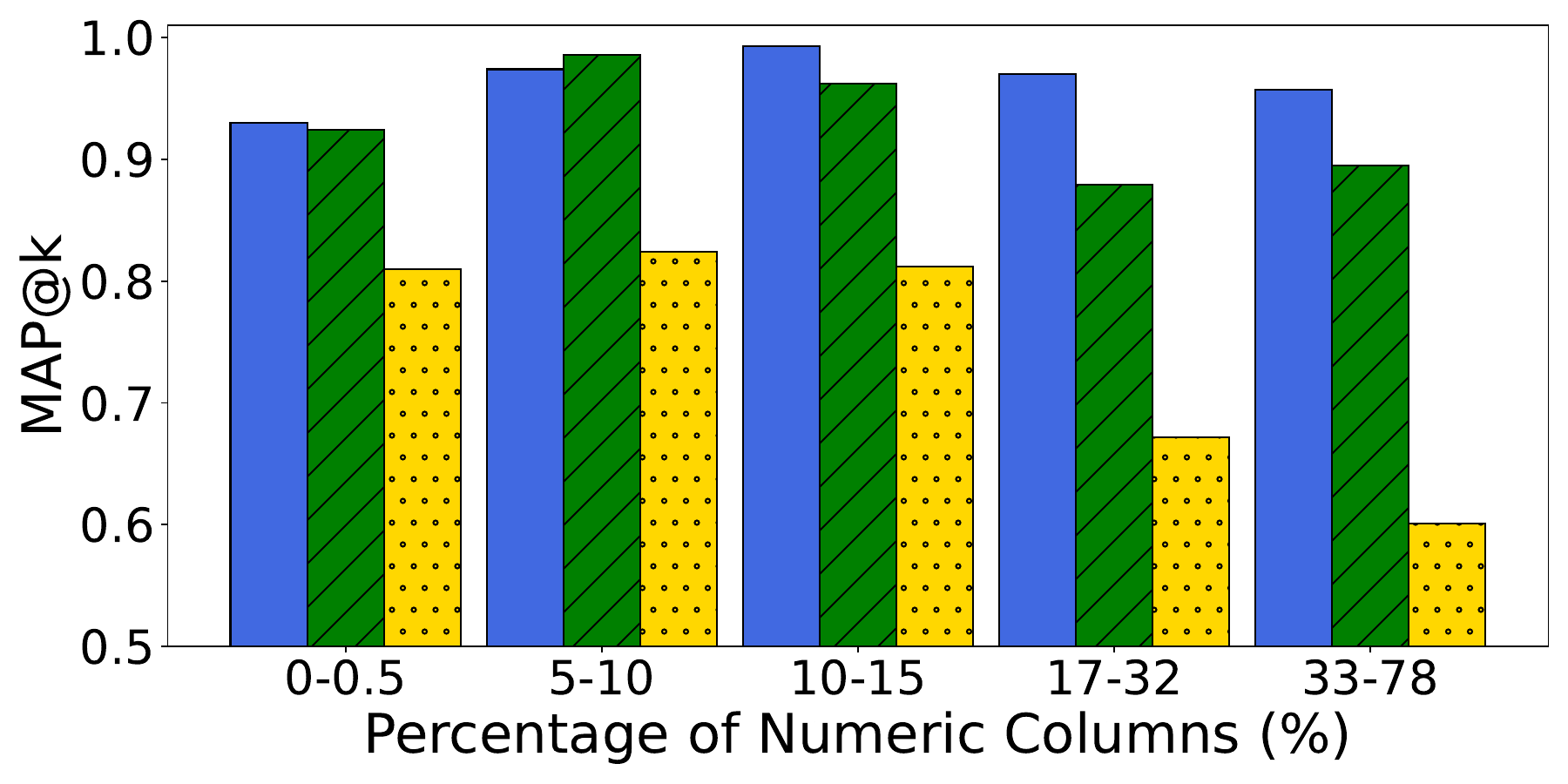}
    \end{minipage}
    }
\caption{In-depth Analyses of \name, SATO, and Sherlock as we vary the number of columns, number of rows, and percentage of numerical columns on the \tusLargeBench benchmark.}
\label{fig:error_tus_large}
\end{figure*}

Conducting the same analyses on the \tusLargeBench benchmark, we see in Figure \ref{fig:error_tus_large} that \name also outperforms the baselines and is consistent across all buckets. On this larger benchmark, even when the baselines' MAP@k performance drops as the number of rows increases or as the percentage of numeric columns increases, \name MAP@k remains consistently high, proving again that \name performs well across tables of various sizes and columns of different types.

\section{Full results for Efficiency experiments}\label{sec-full_efficiency}
\subsection{Efficiency Techniques impact on Performance}
\begin{table}[!ht]
\caption{Efficiency Techniques' impact on query time and performance on the SANTOS labeled benchmark}
\label{tab:perform_runtime}
\small
\begin{tabular}{llcccc}\\ \toprule
Method & Technique & MAP@10 & P@10 & R@10 & Q. Time (sec) \\ \midrule
Starmie & Linear      & 0.993 & 0.984 & 0.737 & 96\\
        & Pruning      & 0.993 & 0.984 & 0.737 & 61\\
        & LSH Index   & 0.932 & 0.780 & 0.580 & 12\\
        & HNSW Index  & 0.945 & 0.810 & 0.606 & 4\\\midrule
SATO    & Linear    & 0.878 & 0.806 & 0.594 & 252\\ 
        & Pruning    & 0.878 & 0.806 & 0.594 & 125 \\
        & LSH Index & 0.818 & 0.712 & 0.528 & 89 \\
        & HNSW Index& 0.730 & 0.520 & 0.378 & 69 \\\midrule
Sherlock& Linear    & 0.782 & 0.672 & 0.493 & 264\\
        & Pruning    & 0.782 & 0.672 & 0.493 & 145 \\
        & LSH Index & 0.737 & 0.612 & 0.449 & 100 \\
        & HNSW Index& 0.705 & 0.550 & 0.406 & 120 \\\midrule
SingleCol& Linear   & 0.891 & 0.798 & 0.588 & 108\\ 
        & Pruning    & 0.891 & 0.798 & 0.588 & 100 \\
        & LSH Index & 0.801 & 0.538 & 0.406 & 11 \\
        & HNSW Index& 0.803 & 0.550 & 0.418 & 2 \\\bottomrule
\end{tabular}
\end{table}

As we experiment with different efficiency techniques in Section \ref{subsec-scal}, we also explore their effects on not only the runtimes but also the effectiveness scores. Table \ref{tab:metrics_runtime} explores the different efficiency techniques for the \name method, showing that they lead to great speedup while preserving the \name performance. In Table \ref{tab:perform_runtime}, we expand on this experiment and apply the efficiency techniques to other embeddings, specifically those of the baselines SATO, Sherlock, and SingleCol. We see that the Pruning technique speeds up the query time by 1.1-2X, while consistently preserving the performance scores perfectly. For indexing techniques, LSH index and HNSW index speed up the query times by 2.6-10X and 2-24X, respectively. Even with the fastest speedup from HNSW index, the baselines SATO and Sherlock are still slower than \name while having worse performance scores. As expected, SingleCol is faster as it does not have the cost from the table context. However, even with the fastest query times from the approximation technique HNSW index, \name still outperforms all baselines.

\subsection{k-Scalability on WDC Benchmark}
\begin{figure}[!ht]
    \centering
    \includegraphics[width=0.3\textwidth]{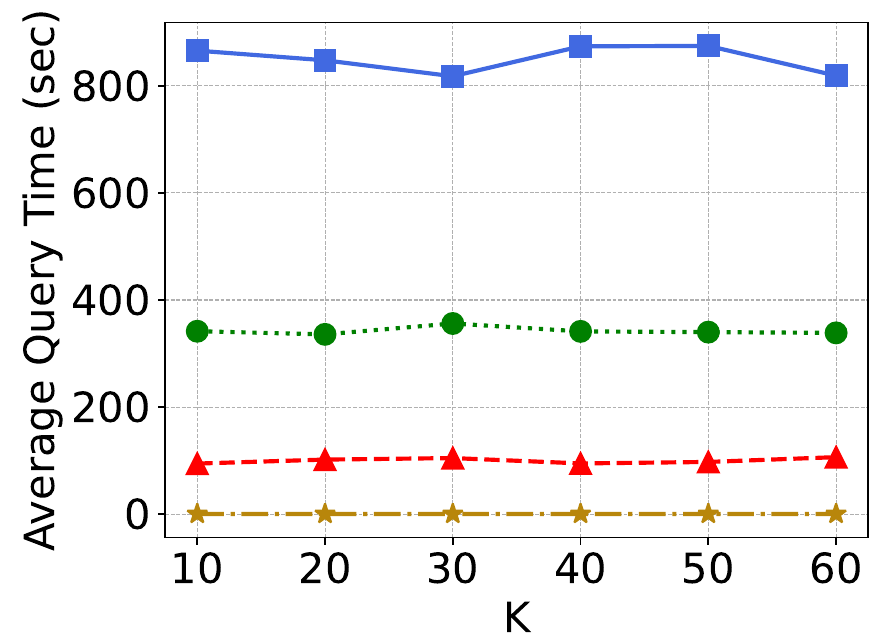}
    \caption{Scalability on 1M WDC tables with varying k's}
    \label{fig:wdc_scal_k}
\end{figure}

For the scalability experiments, in Section \ref{subsec-scal} we experiment with the 4 efficiency techniques on \name on the \realBench and WDC benchmarks. In Figure \ref{fig:scal}(a), we show the experiment on \realBench as we increase k from 10 to 60. In Figure \ref{fig:wdc_scal_k}, we show the same experiment on 1M of the WDC tables. The trends across both figures are similar, with the HNSW index having the fastest query time, followed by LSH index, Pruning, then Linear. However here, HNSW index has a much more impressive performance with the query time remaining around 250 ms as k increases to 60, which is 3000X faster than Linear and 400X faster than LSH index. Thus, \name is generally robust in query time as the number of results to return increases, and is sped up the most with HNSW index.

\section{Discovered Column clusters}\label{app-colcls}

\begin{table*}[!ht]
	\small
	\caption{\small Column clusters discovered by \name. We show the first 3 values from 3 columns 
		of each cluster. The clusters have finer-grained types (e.g., names of schools, grocery stores, song names) than the original ground truth types (e.g., type, name, artist). }\vspace{-1em}
	\label{fig:clusters}
	\begin{tabular}{c|c|c|c}\toprule
		Cluster type        & 1st Column                 & 2nd Column               & 3rd Column             \\ \midrule
		type                   & Emerson Elementary School          & Choctawhatchee Senior High School        & Sumner Academy Of Arts and Science   \\
		→                      & Banneker Elementary School  & Fort Walton Beach High School     & Wyandotte High School          \\
		Names of schools       & Silver City Elementary School  & Ami Kids Emerald Coast             & J C Harmon High School          \\ \midrule
		name                   & People's Grocery Co-op Exchange    & Amazing Grains                &  Apples Street Market\\
		→                      & Prairieland Market                     & BisMan Community Food Cooperative & Bexley Natural Market        \\
		Food/grocery stores    & The Merc (Community Mercantile)               &  Bowdon Locker \& Grocery            & Kent Natural Foods Co-op        \\ \midrule
		artist                 & I Don't Give A ...                 &  Spoken Intro                    &  New Wave                        \\
		→                      & I'm The Kinda                       & The Court                         & Up The Cuts                            \\
		Song names             & I U She                           & Maze                       & Thrash Unreal \\ \bottomrule
	\end{tabular}
\end{table*}

We further inspect the column values within each cluster and find that \name discovers
clusters of finer-grained semantic types not present in the original 78 types. 
Table~\ref{fig:clusters} shows 3 such example clusters. The majority types (from the 78 original types) of columns in the 3 clusters
are ``type'', ``name'', and ``artist'' respectively. After inspecting the column values,
we can interpret the types of the 3 clusters as names of schools, names of food/grocery stores,
and names of songs. 
It is difficult to discover such fine-grained types by existing methods based on supervised classification.

Table \ref{tab:clustering_full} shows the full results
of column clustering. We use Sherlock, Sato, \name,
and its single-column version to generate the column
embeddings. After obtaining the column embeddings,
we construct a similarity graph by adding edges
between pairs of columns with similarity above a threshold
$\tau = 0.6$.
We then cluster the columns by computing their connected 
components. Note that for fair comparison,
we restrict the size of clusters to be around 50
so that different methods generate similar numbers of 
clusters. We measure the quality of clusters by their
purity scores, which measure how likely a column is
assigned to a cluster with the same majority semantic type
as that of the column.
Among the 4 methods, \name generates
clusters with the highest purity score of 51.19\%.

\begin{table}[!ht]
\caption{Purity scores of clusters by \name vs. Sherlock
and Sato.}\label{tab:clustering_full}
\small
\begin{tabular}{cccc} \toprule
                     & n\_clusters & avg. cluster size & Purity (\%) \\ \midrule
Sherlock             & 2,395        & 49.84             & 30.50       \\
Sato                 & 2,456        & 48.60             & 37.36       \\
\name  & 2,297        & 51.96             & 51.19       \\
\name (SingleCol) & 9,252        & 12.90             & 20.38      \\ \bottomrule
\end{tabular}
\end{table}

\section{Full results for data discovery for ML}

Table \ref{tab:ml_full} shows the full results
of the 25 rating prediction tasks created from 
4,130 WDC web tables of $\geq$50 rows.
Each dataset is split into a training and a testing set
at a 4:1 ratio. The baseline methods are:

\smallskip\noindent
\textbf{NoJoin: } Train a XGBoost model with numeric and 
textual features from the original table $S$ only.
We featurize text attributes using 
the Sentence Transformers 
library~\cite{DBLP:conf/emnlp/ReimersG19}.

\smallskip\noindent
\textbf{Jaccard: } Perform an equal left-join with
a table that contains a column with the highest Jaccard
similarity with any column in the query table. Namely,
given a query table $S = \{s_1, \dots, s_n\}$ of $n$ 
non-target columns and a data lake $\bigT$, we join $S$
with the data lake table
$$\argmax_{T \in \bigT}\left(\max_{s_i \in S, t_j \in T}(\mathsf{Jaccard}(s_i, t_j))\right) $$
where $\mathsf{Jaccard}(s_i ,t_j)$ is the token-level Jaccard 
similarity over tokens in query column $s_i$ and 
data lake column $t_j$. Note that we exclude ``rating'' columns
from $T$ to avoid any potential label leakage.

\smallskip\noindent
\textbf{Overlap: } In this baseline, we simply replace Jaccard
similarity from above with the overlap score, i.e., 
$\mathsf{Overlap}(s_i, t_j) := 
|\mathsf{tokens}(s_i) \cap \mathsf{tokens}(t_j)|$.

\smallskip\noindent
\textbf{Starmie: } For \name, we use the learned contextualized
embeddings for measuring similarities of columns.
Since the embeddings capture the table context of each column,
we expect the resulting data tables to be semantically relevant
to the query table. More formally, let $\bigM$ be the learned
column encoder, \name joins $S$ with the table 
\begin{align*}
\argmax_{T \in \bigT}(&\max_{s_i \in S, t_j \in T}(\mathsf{cos}(\bigM(s_i), \bigM(t_j))) + \\
&\max_{t_j \in T}(\mathsf{cos}(\bigM(s_\mathsf{target}), \bigM(t_j))) ) .
\end{align*}
Note that we use the second term with the source
target column $s_\mathsf{target}$ (i.e., ``Rating'') to take into account
the similarity between the target column with columns from the data lake table
$T$.

Lastly, an important implementation detail is to make sure
that the join result has the exact same number of rows with
the query table $S$. This is done by properly left-joining with
the data lake table $T$. This is done via the pandas
DataFrame command:

\begin{verbatim}
  # de-duplicate table T on column t_j
  T = T.drop_duplicates(subset=[t_j]).set_index(t_j)
  # left-join on the column pair (s_i, t_j)
  S.join(T, on=s_i)
\end{verbatim}

\begin{table*}[!ht]
\centering
\caption{Detailed MSE scores of 25 regressions tasks with different data discovery methods. The Reduction columns measure the improvement of each method against NoJoin.}
\label{tab:ml_full}
\begin{tabular}{cccccccc}
\toprule
\#row (train+test) & NoJoin & Jaccard & Reduction & Overlap & Reduction & \name & Reduction \\ \midrule
200                & 0.0820 & 0.0885  & -0.0790   & 0.0862  & -0.0508   & 0.0862                 & -0.0508   \\
200                & 0.2360 & 0.2359  & 0.0003    & 0.2368  & -0.0033   & 0.2359                 & 0.0003    \\
200                & 0.0778 & 0.0653  & 0.1604    & 0.0803  & -0.0316   & 0.0653                 & 0.1604    \\
250                & 0.0008 & 0.0008  & 0.0000    & 0.0008  & 0.0000    & 0.0008                 & 0.0000    \\
644                & 0.0865 & 0.0880  & -0.0174   & 0.0880  & -0.0174   & 0.0880                 & -0.0174   \\
533                & 0.1065 & 0.1235  & -0.1599   & 0.1235  & -0.1599   & 0.1235                 & -0.1599   \\
200                & 0.1269 & 0.1313  & -0.0349   & 0.1223  & 0.0365    & 0.1223                 & 0.0365    \\
200                & 0.0236 & 0.0262  & -0.1080   & 0.0232  & 0.0179    & 0.0262                 & -0.1080   \\
535                & 0.0487 & 0.0409  & 0.1586    & 0.0409  & 0.1586    & 0.0409                 & 0.1586    \\
200                & 0.1598 & 0.1544  & 0.0337    & 0.1195  & 0.2520    & 0.1195                 & 0.2520    \\
200                & 0.0206 & 0.0214  & -0.0389   & 0.0214  & -0.0389   & 0.0214                 & -0.0389   \\
529                & 0.0566 & 0.0441  & 0.2208    & 0.0441  & 0.2208    & 0.0441                 & 0.2208    \\
472                & 0.1731 & 0.1355  & 0.2176    & 0.1355  & 0.2176    & 0.1355                 & 0.2176    \\
200                & 0.0176 & 0.0197  & -0.1178   & 0.0197  & -0.1178   & 0.0192                 & -0.0865   \\
200                & 0.0381 & 0.0350  & 0.0824    & 0.0381  & -0.0001   & 0.0350                 & 0.0824    \\
200                & 0.0118 & 0.0097  & 0.1779    & 0.0101  & 0.1420    & 0.0092                 & 0.2239    \\
200                & 0.0515 & 0.0515  & 0.0000    & 0.0515  & 0.0000    & 0.0515                 & 0.0000    \\
387                & 0.0662 & 0.0655  & 0.0104    & 0.0685  & -0.0344   & 0.0655                 & 0.0104    \\
434                & 0.0988 & 0.0765  & 0.2250    & 0.0765  & 0.2250    & 0.0765                 & 0.2250    \\
200                & 0.0177 & 0.0177  & 0.0018    & 0.0177  & 0.0018    & 0.0177                 & 0.0018    \\
200                & 0.1066 & 0.1066  & 0.0000    & 0.0904  & 0.1522    & 0.0129                 & 0.8790    \\
200                & 0.1064 & 0.0829  & 0.2210    & 0.1026  & 0.0352    & 0.0829                 & 0.2210    \\
200                & 0.1875 & 0.1929  & -0.0285   & 0.1894  & -0.0101   & 0.1894                 & -0.0101   \\
300                & 0.0001 & 0.0001  & 0.1302    & 0.0001  & -0.0222   & 0.0001                 & 0.1302    \\
250                & 0.1488 & 0.1077  & 0.2764    & 0.1152  & 0.2261    & 0.1077                 & 0.2764    \\ \midrule
AVG                & 0.0820 & 0.0753  & 0.0533    & 0.0748  & 0.0480    & 0.0699                 & 0.1050 \\ \bottomrule  
\end{tabular}
\end{table*}

\end{document}